\documentclass[10pt,a4paper,twocolumn,multicol,superscriptaddress,prb,amsmath,amssymb,aps,showpacs]{revtex4}
\usepackage{graphicx}
\usepackage{bm}
\usepackage{dcolumn}
\usepackage{bm}
\usepackage{endnotes}
\usepackage{amsmath,amssymb}
\newcommand{\rr}{{\bf r}}

\newcommand{\nc}{\newcommand}
\nc{\be}{\begin{equation}} \nc{\ee}{\end{equation}}
\nc{\bea}{\begin{eqnarray}} \nc{\eea}{\end{eqnarray}}
\nc{\bean}{\begin{eqnarray*}} \nc{\eean}{\end{eqnarray*}}
\nc{\mb}{\mbox} \nc{\rnc}{\renewcommand} \nc{\Bk}{\mb{\bf k}}
\nc{\vp}{\mb{\bf p}} \nc{\vn}{\mb{\bf n}} \nc{\vq}{\mb{\bf q}}
\nc{\Br}{\mb{\bf r}} \nc{\vz}{\hat {\mb{\bf z}}}
\nc{\vj}{\mb{\boldmath$j$}} \nc{\vg}{\mb{\boldmath$g$}}
\nc{\x}{\mb{\boldmath$x$}} \nc{\A}{\mb{\boldmath$A$}}
\nc{\va}{\mb{\boldmath$a$}} \nc{\vs}{\mb{\boldmath$\sigma$}}
\nc{\vpi}{\mb{\boldmath$\pi$}} \nc{\nab}{\nabla} \nc{\X}{\sf x}
\nc{\Bsigma}{\mb{\boldmath$\sigma$}} \nc{\BB}{\mb{\boldmath$B$}}

\begin{document}


\title{Anomalous Hall effect in a two-dimensional electron gas}        
\author{Tamara S. Nunner}
\affiliation{Institut f\"ur Theoretische Physik, Freie Universit\"at
  Berlin, Arnimallee 14, 14195 Berlin, Germany}
\author{N. A. Sinitsyn}
\affiliation{Department of Physics, Texas A\&M University,
College Station, TX 77843-4242, USA}
\affiliation{CNLS/CCS-3, 
Los Alamos National Laboratory,
Los Alamos, NM 87544, USA}
\author{Mario F. Borunda}
\affiliation{Department of Physics, Texas A\&M University,
College Station, TX 77843-4242, USA}

\author{A. A. Kovalev}
\affiliation{Department of Physics, Texas A\&M University,
College Station, TX 77843-4242, USA}
\author{Ar. Abanov}
\affiliation{Department of Physics, Texas A\&M University,
College Station, TX 77843-4242, USA}
\author{Carsten Timm}
\affiliation{Department of Physics and Astronomy, University of Kansas, Lawrence, KS 66045, USA}
\author{T. Jungwirth}
\affiliation{Institute of Physics  ASCR, Cukrovarnick\'a 10, 162 53
Praha 6, Czech Republic }
\affiliation{School of Physics and Astronomy, University of Nottingham,
Nottingham NG7 2RD, UK}
\author{Jun-ichiro Inoue}
\affiliation{Department of Applied Physics, Nagoya University, Nagoya 464-8603, Japan}
\author{A. H. MacDonald}
\affiliation{Department of Physics, University of Texas at Austin, Austin, Texas 78712-1081, USA}
\author{Jairo Sinova}
\affiliation{Department of Physics, Texas A\&M University,
College Station, TX 77843-4242, USA}
\date{\today}          
\begin{abstract}
The anomalous Hall effect in a magnetic two-dimensional electron gas
with Rashba spin-orbit coupling is studied within the Kubo-Streda
formalism in the presence of pointlike potential impurities. 
We find that all contributions to the anomalous Hall conductivity
vanish to leading order in disorder strength when both chiral subbands are occupied. In the
situation that only the majority subband is occupied, all terms are
finite in the weak scattering limit and the total anomalous Hall
conductivity is dominated by skew scattering. We compare our results
to previous treatments and resolve some of the discrepancies
present in the literature. 
\end{abstract}


\maketitle

\section{Introduction}       

In 1879, Edwin Hall ran a current through a gold foil and discovered
that a transverse voltage was induced when the film was exposed to a
perpendicular magnetic field.~\cite{Hall} The ratio of this Hall voltage to the
current density is the Hall resistivity. For paramagnetic materials, the Hall
resistivity is proportional to the applied magnetic field and Hall
measurements give information about the concentration of free carriers
and determine whether they are holes or electrons. Magnetic films 
exhibit both this ordinary Hall response and an extraordinary or
anomalous Hall response that does not disappear at zero magnetic
field and is proportional to the internal magnetization:
$R_{\rm Hall} = R_o H + R_s M$, where $R_{\rm Hall}$ is the Hall
resistance, $R_o$ and $R_s$ are the ordinary and anomalous Hall
coefficients, $M$ is the magnetization, and $H$ is the applied
magnetic field. The anomalous Hall effect (AHE) is the consequence of
spin-orbit coupling and allows an indirect measurement of the internal magnetization.

Despite the simplicity of the experiment,  the theoretical basis of the AHE is still hotly debated and a source of conflicting reports.~\cite{Sinova:2004_c} 
Different mechanisms  contribute to the  
AHE: an intrinsic mechanism and extrinsic mechanisms such as 
skew-scattering and side-jump contributions. The intrinsic mechanism is
based solely on the topological properties of the Bloch states
originating from the spin-orbit-coupled electronic structure as first
suggested by Karplus and Luttinger.~\cite{Karplus:1954_a}  
Their approach gives an anomalous Hall coefficient $R_s$
proportional to the square of the ordinary resistivity, since the
intrinsic AHE itself is insensitive to impurities. The skew-scattering
mechanism, as first proposed by Smit,~\cite{Smit:1955_a,note_skew} relies on an asymmetric
scattering of the conduction electrons by impurities present in the
material.
Not surprisingly, this skew scattering contribution to $R_s$ is
sensitive to the type and range of the scattering potential and, in
contrast to the intrinsic mechanism, scales linearly with the
diagonal resistivity. The presence of impurities also leads to a
side-step type of scattering, which contributes to a net current
perpendicular to the initial momentum. This is the so-called side-jump
contribution, whose semi-classical interpretation was pointed out by
Berger.~\cite{Berger:1970_a} However, it is not trivial to correctly
account for such contributions in the semiclassical
procedure, making a connection to the microscopic approach very desirable.

The early theories of the AHE involved  complex calculations
with results that where not easy to interpret and often contradicting
each other.~\cite{Nozieres:1973_a} The adversity facing these theories
stems from the origin of the AHE: it appears due to the interband
coherence and not just due to simple changes in the occupation of
Bloch states, as was recognized in the early works of Luttinger and
Kohn.~\cite{Kohn:1957_a,Luttinger:1958_a} Nowadays, most treatments of
the AHE either use the semiclassical Boltzmann transport theory or the
diagrammatic approach based on the Kubo-Streda linear-response
formalism. The equivalence of these two methods for the
two-dimensional Dirac-band graphene system has recently been shown by
Sinitsyn {\it et al.},~\cite{Sinitsyn:2006_b} who explicitly
identified various diagrams of the more systematic Kubo-Streda
treatment with the physically more transparent terms of the
semiclassical Boltzmann approach. 

It is therefore important to also obtain a similarly cohesive understanding
of the AHE in other systems such as  
the two-dimensional (2D) spin-polarized electron gas with Rashba
spin-orbit interaction in the presence of pointlike potential
impurities,  
where a series of previous studies has led to a multitude of
results with discrepancies arising from the focus on
different limits and/or subtle missteps in the
calculations.~\cite{Culcer:2003_a,Dugaev:2005_a, Sinitsyn:2005_a,
  Liu:2005_c, Liu:2006_a, Inoue:2006_a,Onoda:2006_a} It is the purpose
of this paper to review and analyze the previous attempts and to
provide a detailed analysis of all contributions to the AHE in a
two-dimensional electron gas. 
Since we have already demonstrated the equivalence of the Kubo-Streda
formalism and the semiclassical Boltzmann approach with respect to
skew scattering in the two-dimensional electron gas in a previous
paper,~\cite{Borunda:2007_a} we will focus here exclusively on the
diagrammatic formalism  based on the Kubo-Streda treatment. 

The outline of the article is as follows. We start by reviewing and
commenting on previous studies of the AHE in the two-dimensional
electron gas in Sec.~\ref{sec:PreviousTreatments}, where we compare
them with our results and discuss the discrepancies and their
possible origins. In Sec.~\ref{sec:AHEconductivity} we present details
of our calculation within the diagrammatic Kubo-Streda formalism. In
Sec.~\ref{sec:SimpleLimits} we provide simple analytical limits of all
terms of the anomalous Hall conductivity and discuss the full
evaluation in Sec.~\ref{sec:NumericalEvaluation}. Finally, in
Sec.~\ref{sec:conclusion} we present our conclusions.

\section{Comparison with previous approaches}
\label{sec:PreviousTreatments}

Currently there are several publications on the AHE in two dimensional
systems  reaching different quantitative
predictions even in the same limits.~\cite{Culcer:2003_a,Dugaev:2005_a,Sinitsyn:2005_a,Sinitsyn:2006_b,Liu:2005_c,Liu:2006_a,Inoue:2006_a,Onoda:2006_a}
In the present paper we present a
calculation with conclusions that are in disagreement with some
previous studies. On such a background we believe that previous
articles have to be discussed in some details. Below we review the
history of the problem and explain why we think the subject has to be
reconsidered. 

A first study of the AHE in two dimensional systems was done by 
Culcer {\it et al.},~\cite{Culcer:2003_a} who calculated only the intrinsic contribution
to the Hall conductivity for a wide class of two-dimensional systems, including the Rashba two-dimensional electron gas as a special case. 
The intrinsic contribution plays a special role in the theory of the AHE because it is not related to the scattering of electrons but is rather caused by the unusual trajectories of electrons under the action of the electric field. 
However,  the disorder contributions can also be important and further
insight was needed in the quest for a quantitatively rigorous theory
of the dc-AHE.

The first attempts to understand the disorder effects where done independently by two groups,~\cite{Dugaev:2005_a,Sinitsyn:2005_a} each employing different approaches. Dugaev {\it et. al.}~\cite{Dugaev:2005_a}
used the version of the Kubo formula, which expresses the Hall
conductivity in terms of the causal Green functions. The intrinsic
contribution appears as a result of calculations with bare Green
functions, while disorder effects renormalize the quasi-particle life
time and the current vertex. This approach is formally rigorous and is
similar to the one we adopt in our work. However, our final results
are quantitatively different from those found in
Ref.~\onlinecite{Dugaev:2005_a} due to a subtlety in the calculation
of the vertex at the Fermi surface which was
later corrected in the appendix of
Ref. \onlinecite{Sinitsyn:2006_b}. Starting with the equation
for the renormalized vertex $T_x = a k_x + b \sigma_x + c \sigma_y$
and with the assumption that the density of impurities is low, 
they find correctly that $b = 0$ to leading order in $n_i$,
i.e. $a/b\propto n_i$. However, such a term gets multiplied by an
equivalent divergent term within the Kubo
formula leading to a non-zero contribution to the AHE conductivity to
zeroth order in $n_i$.   

In contrast to the previous quantum mechanical approach, Sinitsyn {\it
  et al.}~\cite{Sinitsyn:2005_a} employed the semiclassical wave-packet
approach focusing only on the understanding of the side-jump contribution and
formulating the semi-classical problem in a gauge invariant
form. This work~\cite{Sinitsyn:2005_a} intentionally avoids a
discussion of the skew-scattering contribution due to the asymmetry of
the collision term kernel, which is also an important mechanism of the
Hall current and can even be parametrically similar to all other
contribution~\cite{Sinitsyn:2006_b} in the case of Gaussian
correlations. Therefore, the work in Ref.~\onlinecite{Sinitsyn:2005_a}
is meant as an intuitive introduction into the physics of the
anomalous velocity and the side-jump effect, but  
does not offer a rigorous quantitative comparison even in the
considered limit of smooth disorder potential. 

Subsequently two papers by Liu {\it et al.}~\cite{Liu:2005_c,Liu:2006_a} studied the problem using the Keldysh technique
for linear transport. The Keldysh technique leads to the quantum Boltzmann equation for the diagonal  elements of the density matrix in momentum space when only elastic scattering events are considered. 
In the steady state limit of a weak electric field this equation can be written as follows:
\begin{equation}
e{\bf E \cdot}\nabla_{{\bf p}} \hat{\rho}({\bf p})+
i[\hat{H}_0,\hat{\rho}({\bf p})]=\hat{I}_{\mathrm col}(\hat{\rho}({\bf
  p})) \,,
\label{qbe}
\end{equation}
where $\hat{I}_{\mathrm col}$ contains all disorder dependent terms
that become zero when $\hat{\rho}({\bf p})$ is the density matrix in
thermodynamic equilibrium and $\hat{H}_0$ is the disorder free part of
the Hamiltonian. The ``hat" means that $\hat{\rho}$ and
$\hat{I}_{col}$ are matrices in the band index space. The term
containing the electric field is called the driving term. In the
linear-response approximation it only depends on the equilibrium part of the density matrix.

To start with Eq. (\ref{qbe}) is correct and is also the starting
point of the pioneering work by Luttinger~\cite{Luttinger:1958_a} and
therefore one can compare it directly with steps taken by Liu {\it et
  al.}~\cite{Liu:2005_c,Liu:2006_a} Luttinger's approach was to split
the density matrix into equilibrium and nonequilibrium parts 
$\hat{\rho}=\hat{\rho}_{\mathrm eq}+\hat{\rho}_{\mathrm neq}$
where $\hat{\rho}_{\mathrm neq}$ is linear in electric field. It is
this part of the density matrix that is responsible for nonzero
currents. For weak disorder potential $\hat{V}$, Luttinger looked for
$\hat{\rho}_{\mathrm neq}$ as a series in powers of the disorder
potential. He found that this series starts from the term of the order
$\hat{V}^{-2}$ 
\begin{equation}
\hat{\rho}_{\mathrm neq}=\hat{\rho}_{\mathrm
  neq}^{(-2)}+\hat{\rho}_{\mathrm neq}^{(-1)}+
\hat{\rho}_{\mathrm neq}^{(0)}+\cdots
\label{serrho}
\end{equation} 
As pointed out by Luttinger, the leading order term
$\hat{\rho}_{\mathrm neq}^{(-2)}$ does not contribute to the Hall
effect and is only responsible for the longitudinal diffusive
current. The term $\hat{\rho}_{\mathrm neq}^{(-1)}$ was identified
with skew scattering. This term, however, is parametrically very
distinct and vanishes in the approximation of purely Gaussian
correlations of disorder Fourier components; therefore, Luttinger went
to next order and calculated the term $\hat{\rho}_{\mathrm
  neq}^{(0)}$. He found a number of contributions, whose physical
meaning he did not clarify. The main conclusion was that at this order
both the diagonal and off-diagonal parts of the density matrix become
nonzero and contribute to the Hall conductivity, which  
becomes formally independent on the strength of disorder $\hat{V}$ in
the DC limit, although disorder has to be included in the intermediate
calculations. 

Comparing this with the first work of Liu and Lei~\cite{Liu:2005_c} we
find that they determined self-consistently only the off-diagonal part
of the density matrix in band index. This is, however, not enough for
a rigorous quantitative result because the diagonal part of the
$\hat{\rho}_{neq}^{(0)}$ contribution has been known to be important
since Luttinger's pioneering work.  

In their next effort Liu {\it et al.}~\cite{Liu:2006_a} studied the problem of 2D Rashba systems in small gap semiconductor materials, in which a projection to the conduction band leads to extrinsic type spin-dependent contributions. In this work they noticed that the diagonal part is important and calculated it numerically. 
For the driving term in Eq.~(\ref{qbe}) Liu {\it et al.} assume that $\hat{\rho}_{eq}$ is just a diagonal equilibrium Fermi
distribution. This would be correct if one was using the basis of the eigenstates of the {\em full} Hamiltonian with impurities. 
However, both Liu {\it et al.} and Luttinger work in the chiral basis of the disorder free Hamiltonian $\hat{H}_0$. In this basis the {\em equilibrium} state density matrix is no longer diagonal and can also be written as a series in powers of the disorder potential:
\begin{equation}
\hat{\rho}_{eq}=\hat{\rho}_{eq}^{(0)}+\hat{\rho}_{eq}^{(2)}+\cdots
\label{reqe}
\end{equation}
Luttinger has shown that in order to properly evaluate the non-equilibrium part $\hat{\rho}_{neq}^{(0)} $ one should include the second term $\hat{\rho}_{eq}^{(2)}$ of the expansion of the equilibrium density matrix in Eq.~(\ref{reqe}) into the driving term of Eq.~(\ref{qbe}). This was not done in Ref.~\onlinecite{Liu:2006_a} and therefore we believe that their work is incomplete due to such omission. We also note that the correction of order $\hat{V}^2$ in Eq.~(\ref{reqe})   leads to the Hall current contribution, which was identified in the semiclassical approach~\cite{Sinitsyn:2006_a} as the anomalous distribution correction and if omitted leads to errors of factors of two in the typical side-jump type contributions.\cite{Nozieres:1973_a} In the Kubo formula approach, neglecting this correction would be equivalent to the unjustified omission of an important subset of Feynman diagrams.~\cite{Sinitsyn:2006_b}  Within the calculation presented here all these terms are present.

Inoue {\it et al.}~\cite{Inoue:2006_a} calculated the AHE contribution using the same approach we use focusing on the limit of both subbands being occupied and, in addition to the disorder that we consider, incorporating magnetic impurities in the model Hamiltonian. They found that for paramagnetic impurities the Hall conductivity vanishes. Our more general calculations confirm this result. However, we point to one
important difference in its derivation. In both cases the dc-limit Kubo formula, where the conductivity is expressed via
retarded and advanced Greens functions, has been employed to calculate the Hall conductivity. As was shown by Streda ~\cite{Streda:1982_a},  this version of the Kubo formula contains two parts: $\sigma^I_{xy}$ a contribution from the Fermi
surface  and $\sigma^{II}_{xy}$ a contribution from all states of the Fermi sea.
The latter part is less known because it does not appear in the
expression for the longitudinal conductivity. Inoue {\it et al.}
\cite{Inoue:2006_a} calculated only $\sigma^I_{xy}$ and indeed we find
that for their choice of parameters the second part of the
conductivity $\sigma^{II}_{xy}$ vanishes, explaining the agreement
with our results. In a more general analysis, beyond the limit of weak
spin-orbit and Zeeman couplings, we find a non-vanishing
$\sigma^{II}_{xy}$. Our work provides the missing estimate of
$\sigma^{II}_{xy}$ and extends the calculations of Inoue {\it et
  al.}~\cite{Inoue:2006_a}

Finally, the latest work on the subject is by Onoda {\it et
  al.}~\cite{Onoda:2006_a} The authors used the Keldysh
technique, which they reformulated in a way appropriate for multiband
problems in a gauge invariant formalism. They also derived a
self-consistent equation, which is the analog of the standard quantum
Boltzmann equation and solved it numerically. Unfortunately,  lacking
a full understanding of the details of the numerical procedure
and the starting equations being very formal within a
non-chiral basis, a detailed discussion of their approach cannot be
performed here. However, being devoted to the same model, the final
results can be compared directly with the possible discrepancies
arising from the different limits considered in the disorder
distributions in which $n_i$ and the disorder strength are two
independent parameters in their calculations. Onoda {\it et
  al.}~\cite{Onoda:2006_a} find a strong 
skew scattering contribution of the order of $\epsilon_{SO}V_{\mathrm imp}
\sigma_{xx}/W^2$, where $W$ is the inverse density of states. The skew
scattering term changes sign at the point where the minority band
becomes depleted, which they call the resonance point. 
The authors find also that the side-jump contribution is small in
comparison with the intrinsic one. Our results confirm
neither of those predictions. We find that for the Rashba model with 
randomly placed delta-function impurities the leading part of the
skew-scattering vanishes identically when the Fermi level is above
this resonance point. Although skew scattering could still appear in
higher order terms of the Born series, we expect these contributions
to be small because they are of higher order in $V_{\mathrm imp}$. On the
other hand, Onoda {\it et al}~\cite{Onoda:2006_a} consider the limit
of dilute impurities $n_i\rightarrow 0$ independently of the disorder
strength $V_{\mathrm imp}$ which might be the origin for the discrepancies. Using
the Keldysh formalism in the disorder free basis we have been able to
verify analytically our results. Further numerical analysis~\cite{Kovalev:2007_b} of 
different limits  will be necessary to settle the
discrepancies with the results by Onoda {\it et
  al}.~\cite{Onoda:2006_a}

\section{Anomalous Hall conductivity of the 2DEG}
\label{sec:AHEconductivity}

\subsection{Model Hamiltonian}

\begin{figure}[t]
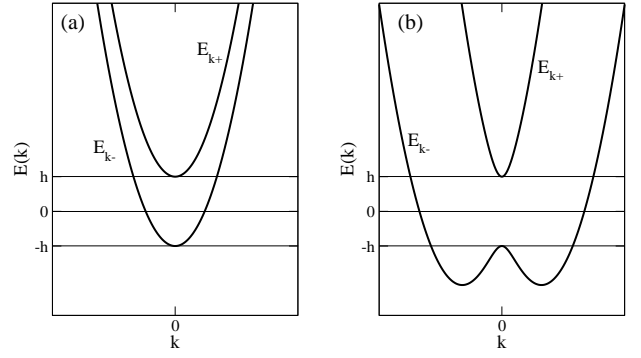

\begin{flushleft}
\begin{minipage}{.49\columnwidth}
\includegraphics[width=.9\columnwidth,clip]{Disp_alpha02.eps}
\end{minipage}
\begin{minipage}{.49\columnwidth}
\includegraphics[width=.9\columnwidth,clip]{Disp_alpha20.eps}
\end{minipage}
\end{flushleft}
\caption{Single particle dispersion for small spin-orbit interaction
  $\alpha k_F/h=0.2$ (a) and large spin-orbit interaction $\alpha
  k_F/h=2.0$ (b).}  
\label{fig:disp}
\end{figure}

We consider a spin-polarized two dimensional electron gas with Rashba spin-orbit interaction
\be \label{eq:Hamiltonian}
H = \frac{k^2}{2m} \sigma_0 + \alpha (\sigma_x k_y - \sigma_y
k_x) - h \sigma_z +  V(\rr)\sigma_0 
\ee
where $m$ is the the effective in-plane mass of the quasiparticles, $\alpha$ the spin-orbit coupling parameter, $h$ the exchange field, and $\sigma_i$ the $2 \times 2$ Pauli matrices. The eigenenergies of the clean system are 
\begin{equation}
E_{k\pm} = \frac{k^2}{2m} \pm \lambda_k \quad {\rm with} \quad
\lambda_k=\sqrt{h^2 + \alpha^2 k^2}  
\label{eq:eigenenergies}
\end{equation}
and are shown in Fig.~\ref{fig:disp}.
The retarded Greens function of the clean system is:
\begin{eqnarray}
G^{(0)R}&=&\frac{\left(\omega-\frac{k^2}{2m} +i 0^+ \right) \sigma_0
       + \alpha k_y \sigma_x - \alpha k_x \sigma_y 
       - h \sigma_z}
{\left(\omega-\frac{k^2}{2m} +i
    0^+ \right)^2-h^2-\alpha^2 k^2} \nonumber \\
&=& G^{(0)R}_0 \sigma_0 + G^{(0)R}_x \sigma_x + G^{(0)R}_y \sigma_y + G^{(0)R}_z \sigma_z \,,
\label{eq:G0}
\end{eqnarray}
with
\begin{eqnarray}
G^{(0)R}_0 \!\!\! &=&\!\!\! \frac{1}{2} ( G^{(0)}_+ \!\!+ G^{(0)}_- ) \,\,\,\,\,
G^{(0)R}_z \!=\! -\frac{1}{2} 
\frac{h}{\lambda_k}
(G^{(0)}_+ \!\!- G^{(0)}_-) \\
G^{(0)R}_x \!\!\!&=& \!\!\! \frac{1}{2} 
\frac{\alpha k_y}{\lambda_k}
(G^{(0)}_+ \!\!- G^{(0)}_-) \,\,\,\,\,
G^{(0)R}_y \!=\! - \frac{1}{2} 
\frac{\alpha k_x}{\lambda_k}
(G^{(0)}_+ \!\!- G^{(0)}_-)  \nonumber
\end{eqnarray}
and
\begin{equation}
G^{(0)}_\pm=\frac{1}{\omega-E_{k\pm}+i0^+} \,.
\end{equation}
The disorder potential $V(\Br)$ in Eq.~(\ref{eq:Hamiltonian}) is
assumed as spin-independent. We consider the model of randomly located
$\delta$-function scatterers: $V({\bf r}) =\sum_i V_i \delta ({\bf
  r}-{\bf R}_i)$ with random and strength distributions
satisfying $\langle V_i \rangle_{\mathrm dis} =0$, $\langle V_i^2
\rangle_{\mathrm dis} =V_0^2 \ne 0$ and $\langle V_i^3
\rangle_{\mathrm dis} = V_1^3 \ne 0$. This model is different from the
standard white noise disorder 
model in which only the second order cumulant is nonzero; $\langle
|V^0_{{\bf k'k}}|^2 \rangle_{dis} = n_i V_0^2 $ where $n_i$ is the
impurity concentration and other correlators are either zero or
related to this correlator by Wick's theorem. The deviation from white
noise in our model is quantified by $V_1 \neq 0$ and is necessary to
capture part of the skew scattering contribution to the anomalous Hall
effect.  
 
We calculate the self-energy using the Born approximation:
\begin{eqnarray}
\label{eq:SelfEnergy}
\Sigma^R&=& -i (\Gamma \sigma_0 + \Gamma_z \sigma_z)\\
&=& \!\!\!- \frac{i}{4} n_i 
V_0^2   \left( (\nu_+ + \nu_-) \sigma_0
- h \left( \frac{\nu_+}{\lambda_+} - \frac{\nu_-}{\lambda_-} \right)
\sigma_z \right) 
\nonumber
\end{eqnarray}
where $\nu_\pm$ is related to the density of states at the Fermi levels of the two subbands
\begin{equation}
\nu_\pm = k \left | \frac{d E_{k\pm}}{dk} \right|^{-1}
=\frac{m \lambda_\pm}{\sqrt{\lambda_F^2+(\alpha^2 m)^2}}
\label{eq:nu}
\end{equation}
with
\begin{equation}
\lambda_\pm=\sqrt{h^2+\alpha^2 k_\pm^2}=
\sqrt{\lambda_F^2+(\alpha^2 m)^2} \mp \alpha^2 m \,, \quad
\end{equation}
where $\lambda_F=\sqrt{h^2+2 \alpha^2m\epsilon_F}$
and 
\begin{equation}
k_\pm=\sqrt{2m \left (\epsilon_F + \alpha^2 m \mp 
\sqrt{\lambda_F^2+(\alpha^2 m)^2} \right)} 
\end{equation}
are the Fermi momenta of the two subbands.

Including the self-energy, the impurity averaged Greens function becomes:
\begin{eqnarray}
G^R \!\!\!&=& \!\!\frac{\!\left(\omega-\frac{k^2}{2m} +i \Gamma\right) \sigma_0
       + \alpha k_y \sigma_x - \alpha k_x \sigma_y 
       - (h+i\Gamma_z) \sigma_z}
{\left(\omega-\frac{k^2}{2m} +i
    \Gamma\right)^2-(h+i\Gamma_z)^2-\alpha^2 k^2} \nonumber \\
&=& \!\! G^R_0 \sigma_0 + G^R_x \sigma_x + G^R_y \sigma_y + G^R_z \sigma_z \,.
\end{eqnarray}
By comparing this expression with Eq.~(\ref{eq:G0}) one observes that the
impurity averaged Greens function can be obtained from the Greens
function of the clean system by the following replacements:
\begin{equation}
\omega \to \omega + i \Gamma \,, \quad
h \to h + i \Gamma_z \,.
\end{equation}
In the limit of small $\Gamma_z$ one can therefore expand
\begin{equation}
\lambda_k \to \sqrt{(h+i \Gamma_z)^2 + \alpha^2 k^2}
\approx \lambda_k \left( 1+ i \frac{h \Gamma_z}{\lambda_k^2} \right) \,.
\end{equation}
Using this approximation the impurity averaged Greens function can
also be written as:
\begin{eqnarray}
\label{eq:GreensFunction}
G^R_0 &=& \frac{1}{2} ( G^R_+ + G^R_- ) \\
G^R_x &=& \sin \phi \, \tilde G^R_x = \frac{1}{2} 
\frac{\alpha k_y \lambda_k}{\lambda_k^2 + i\Gamma_z h}
(G^R_+ - G^R_-) 
\nonumber \\
G^R_y &=& \cos \phi \, \tilde G^R_y = - \frac{1}{2} 
\frac{\alpha k_x \lambda_k}{\lambda_k^2 + i\Gamma_z h}
(G^R_+ - G^R_-) 
\nonumber \\
G^R_z &=& -\frac{1}{2} 
\frac{\lambda_k(h+i\Gamma_z)}{\lambda_k^2+i\Gamma_z h}
(G^R_+ - G^R_-)
\nonumber
\end{eqnarray}
with
\begin{equation}
G^R_\pm=\frac{1}{\omega-E_{k\pm}+i\Gamma_{\pm}} 
\end{equation}
and
\begin{equation}
\Gamma_\pm= \Gamma \mp \Gamma_z \frac{h}{\lambda_\pm} \,.
\end{equation}

\subsection{General expression for the anomalous Hall conductivity}

According to the Kubo-Streda formalism~\cite{Streda:1982_a} the off-diagonal conductivity can be written as:
\begin{equation}
\sigma_{yx}=\sigma_{yx}^{I(a)}+\sigma_{yx}^{I(b)}+\sigma_{yx}^{II}
\label{eq:TotAHE}
\end{equation}
where
\begin{eqnarray}
\label{eq:sigmaI_II}
\sigma_{yx}^{I(a)} \!\!&=&\!\! \frac{e^2}{2\pi V} {\rm Tr} \langle
v_y G^R(\epsilon_F) v_x G^A(\epsilon_F) \rangle
\\
\sigma_{yx}^{I(b)}\!\! &=& \!\!-\frac{e^2}{4\pi V} {\rm Tr} \langle
v_y G^R(\epsilon_F) v_x G^R(\epsilon_F) \nonumber \\
&&  \hspace{1.2cm} + v_y G^A(\epsilon_F) v_x G^A(\epsilon_F) \rangle
\nonumber \\
\sigma_{yx}^{II} \!\!&=& \!\!\frac{e^2}{4\pi V}  
\int_{-\infty}^{\infty} d\epsilon f(\epsilon) {\rm Tr} \langle
v_y G^R(\epsilon) v_x \frac{\partial G^R(\epsilon)}{\partial
  \epsilon}\nonumber \\
&&-v_y \frac{\partial G^R(\epsilon)}{\partial \epsilon} v_x G^R(\epsilon)
-v_y G^A(\epsilon) v_x \frac{\partial G^A(\epsilon)}{\partial
  \epsilon}\nonumber \\
&&+v_y \frac{\partial G^A(\epsilon)}{\partial \epsilon} v_x G^A(\epsilon)
\rangle \,. \nonumber 
\end{eqnarray}
Here, $\sigma^{I}$ results from the electrons at the Fermi surface
whereas $\sigma^{II}$ denotes the contribution of all states of the
Fermi sea. For $\sigma^{I(b)}$ and $\sigma^{II}$ it is sufficient to
calculate the bare bubble contribution in the weak scattering
limit~\cite{Sinitsyn:2006_b} because vertex corrections are of higher
order in the scattering rate $\Gamma$. Plugging in the Greens function
of Eq.~(\ref{eq:GreensFunction}) and using the velocity vertices 
\begin{equation}
v_x=\frac{k_x}{m} \sigma_0 - \alpha \sigma_y \,, \quad
v_y=\frac{k_y}{m} \sigma_0 + \alpha \sigma_x \,
\label{eq:velocities}
\end{equation}
one finds that $\sigma^{I(b)}$ vanishes
\begin{eqnarray}
\sigma_{yx}^{I(b)}&=&-\frac{e^2}{4 \pi V} \frac{1}{(2 \pi)^2} \int d^2 k 
\left ( -i \alpha^2 G_0^R G_z^R +i \alpha^2 G_z^R G_0^R \nonumber
\right. \\
&&        \left. -i \alpha^2 G_0^A G_z^A +i \alpha^2 G_z^A G_0^A \right)
=0 \,.
\end{eqnarray}
The bare contribution of $\sigma^{II}$ in the clean limit, i.e., for
$\Gamma_+=\Gamma_-=0^+$ can be
calculated by integration (see App.~\ref{app:IntSigmaII}) and yields
\begin{eqnarray}
\sigma_{yx}^{II} 
\!=\! \frac{e^2}{4 \pi} \!\left (\!\! 1 \!-\!
\frac{h}{\sqrt{h^2 + 2 \alpha^2 m \epsilon_F +(\alpha^2m)^2}} \! \right)
\!\! \Theta(h-\epsilon_F) 
\label{eq:sigma2}
\end{eqnarray}
where $\partial G_\pm^{R/A}/\partial \epsilon=-(G_\pm^{R/A})^2$ has been used. Including the real scattering rates $\Gamma_+$ and $\Gamma_-$ does not lead to qualitatively different results but mainly causes a slight smearing. Thus we consider it as sufficient to focus on the clean limit contribution of $\sigma^{II}$.

\begin{figure}[t]
\begin{flushleft}
\includegraphics[width=.4\columnwidth]{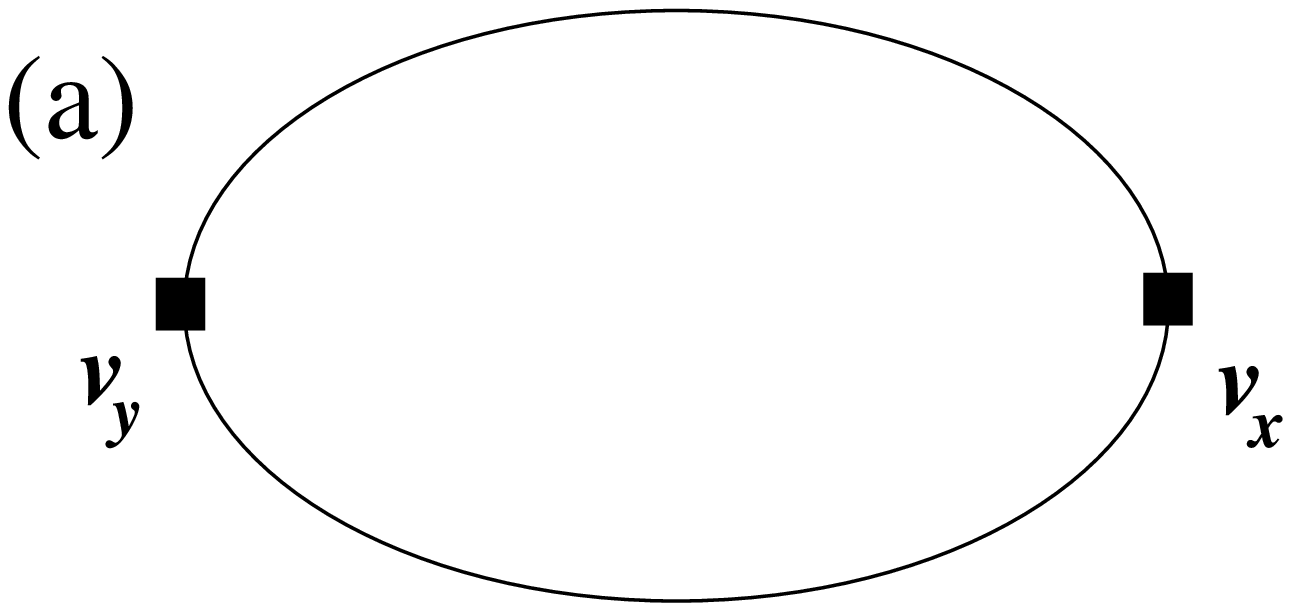}\\[.5cm]
\hspace{-.5cm}
\begin{minipage}{.49\columnwidth}
\includegraphics[width=.85\columnwidth]{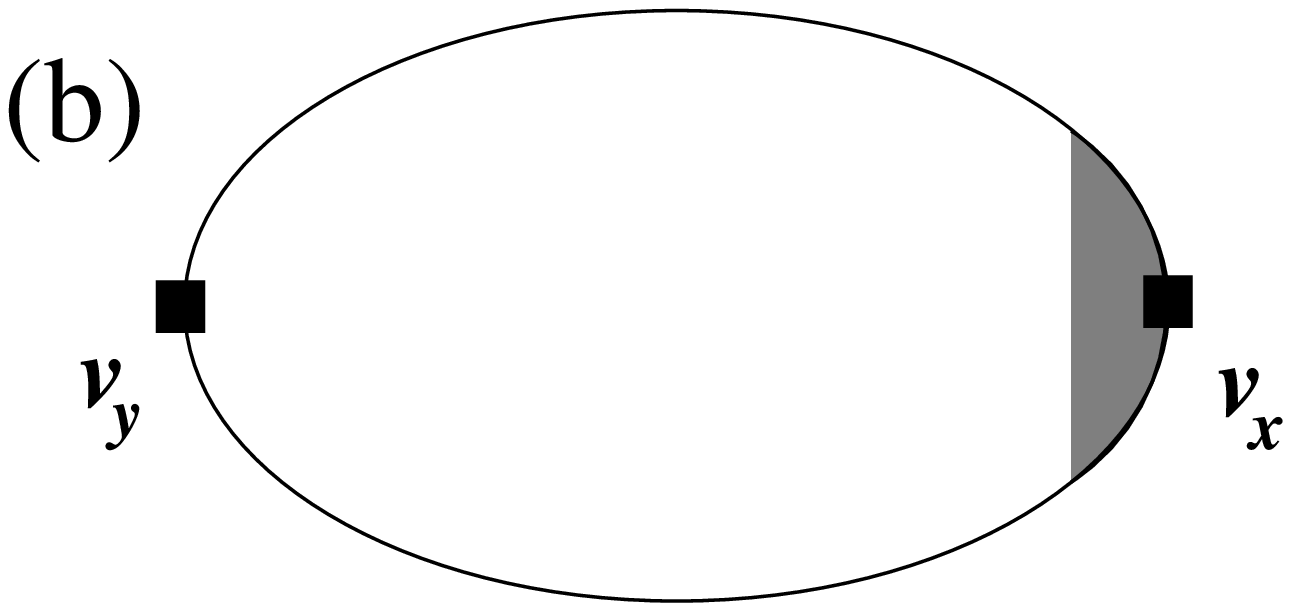}
\end{minipage}
\begin{minipage}{.49\columnwidth}
\includegraphics[width=.85\columnwidth]{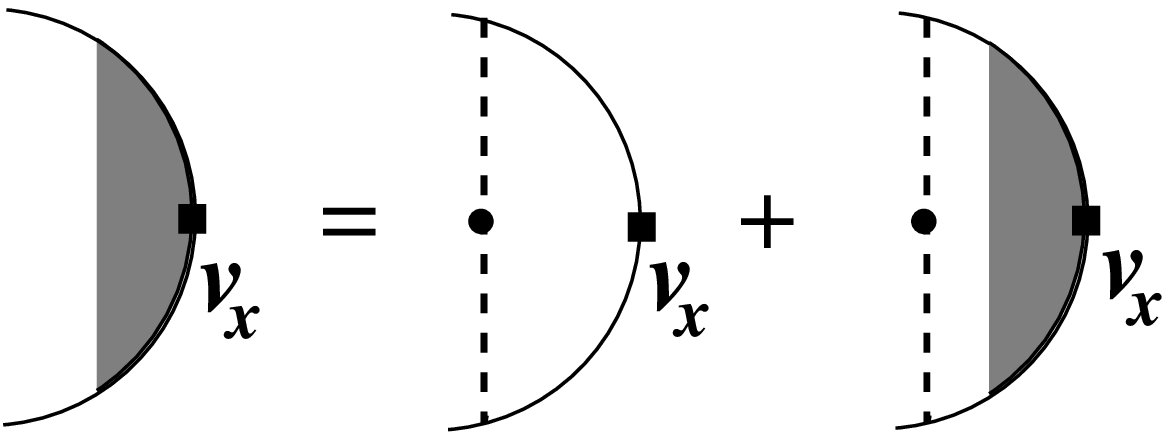}
\end{minipage}\\[.5cm]
\includegraphics[width=.9\columnwidth]{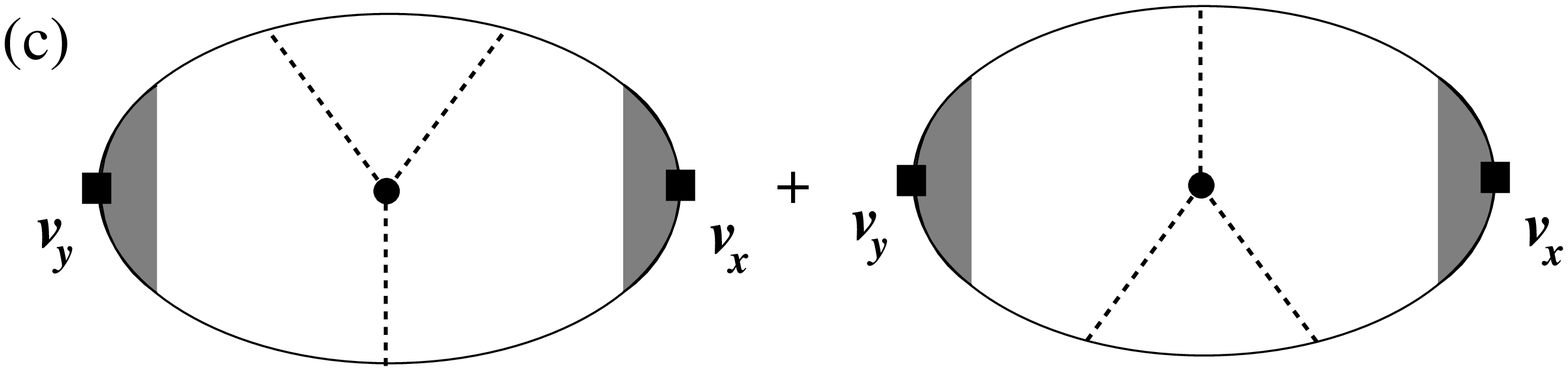}
\end{flushleft}
\caption{Diagrammatic representation of the bare bubble (a), of the ladder vertex
  corrections (b) and of the skew scattering contribution (c).}  
\label{fig:SigmaIa}
\end{figure}

For $\sigma^{I(a)}$ vertex corrections can be of similar magnitude as the bare bubble and thus have to be considered carefully. In the weak scattering limit contributions of higher order impurity scattering vertices are small leaving only ladder type vertex corrections and the $V_1^3/(n_i V_0^4)$ skew scattering contribution as the important terms.~\cite{Borunda:2007_a} Thus we decompose $\sigma^{I(a)}$ in the following way:
\begin{equation}
\sigma_{yx}^{I(a)}=\sigma_{yx}^{I(a),b}
+ \sigma_{yx}^{I(a),l} + \sigma_{yx}^{I(a),s}
\,,
\end{equation}
where $\sigma_{yx}^{I(a),b}$ is the bare bubble contribution (Fig.~\ref{fig:SigmaIa}(a)), $\sigma_{yx}^{I(a),l}$ the ladder vertex corrections (Fig.~\ref{fig:SigmaIa}(b)), and $\sigma_{yx}^{I(a),s}$ the skew scattering contribution (Fig.~\ref{fig:SigmaIa}(c)). With respect to the skew scattering contribution we have shown~\cite{Borunda:2007_a} that only the diagrams with a single third order vertex (see Fig.~\ref{fig:SigmaIa}(c)) contribute to order
$V_1^3/(n_i V_0^4)$. In this diagram both vertices have to be renormalized by ladder vertex corrections. 

\subsubsection{Bare bubble}

The calculation of the bare bubble contribution proceeds as follows: 
\begin{eqnarray}
\label{eq:BareBubble}
&&\sigma_{yx}^{I(a),b}=
\frac{e^2}{2\pi} 
\!\!\int\!\!\! \int \frac{dk k d\phi}{(2 \pi)^2}
{\rm Tr} [v_y G^R(\epsilon_F) v_x G^A(\epsilon_F) ] 
\nonumber \\
&&= 2i\alpha \!\! \int \!\! \frac{dk k}{2\pi}
\left( \frac{k}{m} (\tilde G_y^R G_z^A - G_z^R \tilde G_y^A)
-\alpha  (G_0^R G_z^A - G_z^R G_0^A)\right) \nonumber \\
&&=2 i \alpha (2 I_3 - \alpha I_2)
\end{eqnarray}
where (for explicit evaluation of integrals $I_1$, $I_2$, $I_3$ and $I_4$
see App.~\ref{app:Integrals})
\begin{eqnarray}
\label{eq:integrals}
&&I_1 = \frac{1}{2\pi} \int dk k \left (G_0^R G_0^A - G_z^R G_z^A
\right) \\
&&\approx \frac{1}{8} \left (
          \left(1-\frac{h^2}{\lambda_+^2} \right) \frac{\nu_+}{\Gamma_+}
          +\left(1-\frac{h^2}{\lambda_-^2} \right) \frac{\nu_-}{\Gamma_-} 
\right ) \nonumber \\
&&I_2 = \frac{1}{2\pi} \int dk k \left (G_0^R G_z^A - G_z^R G_0^A \right)
\nonumber \\
&& \approx - \frac{i}{4} \left ( 
\frac{\nu_+ h}{\lambda_+^2} + \frac{\nu_- h}{\lambda_-^2}
-\frac{\Gamma_z}{\Gamma_+} \frac{\nu_+ \alpha^2 k_+^2}{\lambda_+^3}  
+\frac{\Gamma_z}{\Gamma_-} \frac{\nu_- \alpha^2 k_-^2}{\lambda_-^3}  
\right ) \nonumber \\
&&I_3 = \frac{1}{2\pi} \int dk \frac{k^2}{2m} 
\left (\tilde G_y^R G_z^A - G_z^R \tilde G_y^A \right)
\nonumber \\
&&\approx -\frac{i}{4} \alpha \Gamma_z \left (
    \frac{\nu_+}{\Gamma_+ \lambda_+} \left (
    \frac{\epsilon_F}{\lambda_+} -1 \right)
   + \frac{\nu_-}{\Gamma_- \lambda_-} \left (
    \frac{\epsilon_F}{\lambda_-} +1 \right)
\right ) \nonumber \\
&&I_4 = \frac{1}{2\pi} \int dk \frac{k^2}{2m} 
\left (G_0^R \tilde G_y^A + \tilde G_y^R G_0^A \right)
\nonumber \\
&&\approx - \frac{1}{4} \alpha \left (
\epsilon_F \left( \frac{\nu_+}{\Gamma_+ \lambda_+}
                 -\frac{\nu_-}{\Gamma_- \lambda_-} \right )
-\left( \frac{\nu_+}{\Gamma_+}+\frac{\nu_-}{\Gamma_-} \right) 
\right ) \,. \nonumber
\end{eqnarray}

\subsubsection{Ladder diagrams}
For the ladder terms $\sigma_{yx}^{I(a),l}$ we sum the vertex corrections in front of the $v_x$ vertex as indicated in Fig.~\ref{fig:SigmaIa}(b). Starting from the momentum integrated bare velocity vertex
\begin{equation}
\!\!\int\!\!\! \int \frac{dk k d\phi}{(2 \pi)^2}
G^R(\epsilon_F) v_x G^A(\epsilon_F) =
\gamma_x \sigma_x + \gamma_y \sigma_y \,,
\end{equation}
with
\begin{equation}
\gamma_x = i (I_3 -\alpha I_2) \,, \quad
\gamma_y = I_4 -\alpha I_1
\label{eq:BareVertices}
\end{equation}
one finds for the renormalized vertex 
\begin{eqnarray}
\label{eq:RenVertex}
\Gamma_{v_x}&=&\Gamma_x \sigma_x + \Gamma_y \sigma_y\\
&=& \gamma_x \sigma_x + \gamma_y \sigma_y \nonumber \\
&&+ n_i V_0^2 \!\!\int\!\!\! \int \frac{dk k d\phi}{(2 \pi)^2}
  G^R(\epsilon_F) (\gamma_x \sigma_x + \gamma_y \sigma_y)
  G^A(\epsilon_F) \nonumber \\
& =&\gamma_x \sigma_x + \gamma_y \sigma_y \nonumber \\
&&+ n_i V_0^2 \left( (I_1 \Gamma_x + i I_2 \Gamma_y) \sigma_x
           + (I_1 \Gamma_y - i I_2 \Gamma_x) \sigma_y \right)
\nonumber
\end{eqnarray}
and thus
\begin{eqnarray}
\label{eq:vertexcorrections}
\begin{pmatrix}
\Gamma_x \\ \Gamma_y
\end{pmatrix}
&=& \frac{1}{(1-n_i V_0^2 I_1)^2-(n_i V_0^2 I_2)^2}\\
&&\begin{pmatrix}
1-n_i V_0^2 I_1 & i n_i V_0^2 I_2 \\ - i n_i V_0^2 I_2 & 1-n_i V_0^2 I_1
\end{pmatrix}
\begin{pmatrix}
\gamma_x \\ \gamma_y
\end{pmatrix} \,. \nonumber
\end{eqnarray}
The ladder diagrams are therefore given by
\begin{eqnarray}
\label{eq:LadderDiagrams}
\sigma_{yx}^{I(a),l} \!\!\!&=&\!\!\! \frac{e^2}{2\pi}
\!\!\int\!\!\!\!\! \int \!\frac{dk k d\phi}{(2 \pi)^2}
{\rm Tr} [G^A \!(\epsilon_F) v_y G^R \!(\epsilon_F) 
(\Gamma_x \sigma_x \!+\! \Gamma_y \sigma_y)] 
\nonumber \\
\!\!\!&=&\!\!\! - \frac{e^2}{2\pi} 2
(\gamma_y \Gamma_x + \gamma_x \Gamma_y)\\
\!\!\!&=&\!\!\! -\frac{e^2}{\pi}
\frac{n_i V_0^2
\!\left( 2 \gamma_x \gamma_y (1\!\!-\!\!n_i V_0^2 I_1)
\!\!+\! i n_i V_0^2 I_2 (\gamma_y^2 \!\!-\!\! \gamma_x^2) \right)}
{(1 -  n_i V_0^2 I_1)^2-(n_i V_0^2 I_2)^2} \,.
\nonumber
\end{eqnarray}
In the weak scattering limit this reduces to
\begin{equation}
\sigma_{yx}^{I(a),l} =-\frac{e^2}{\pi}
\frac{n_i V_0^2
\!\left( 2 \gamma_x \gamma_y (1\!\!-\!\!n_i V_0^2 I_1)
\!\!+\! i n_i V_0^2 I_2 \gamma_y^2  \right)}
{(1 -  n_i V_0^2 I_1)^2} \,.
\end{equation}

\subsubsection{Skew scattering}
\begin{figure}[t]
\begin{center}
\includegraphics[width=.85\columnwidth]{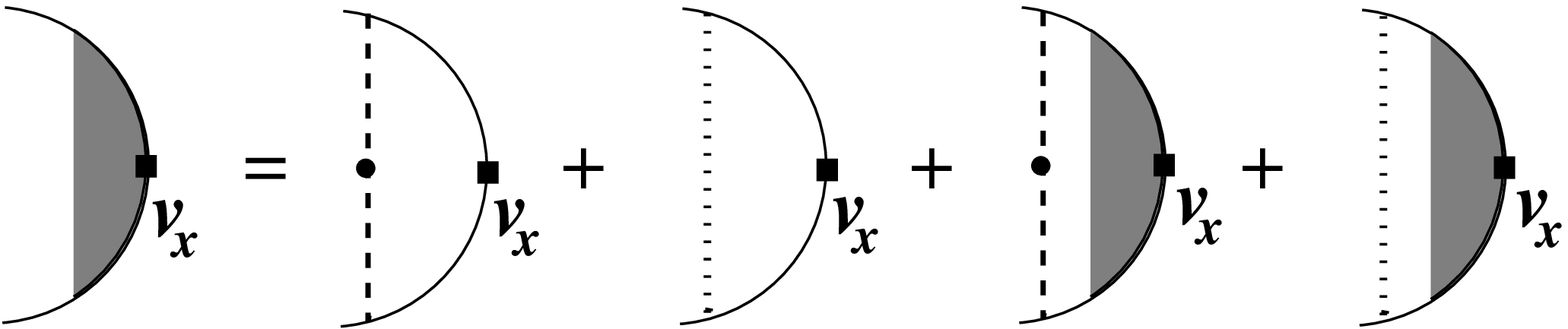}\\[.5cm]
\begin{minipage}{.2\columnwidth}
\hspace*{1cm} with
\end{minipage}
\begin{minipage}{.78\columnwidth}
\hspace*{-1cm}
\includegraphics[width=.6\columnwidth]{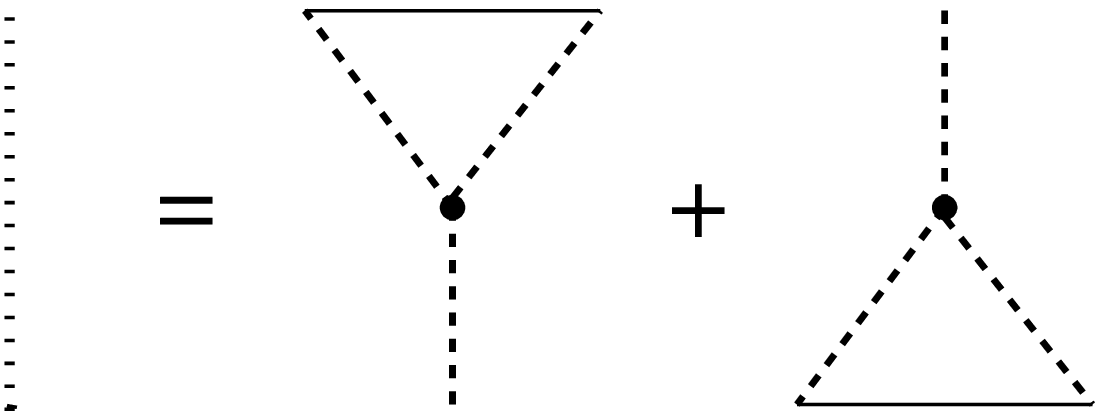}
\end{minipage}
\end{center}
\caption{Full vertex including ladder and skew scattering diagrams.}  
\label{fig:SkewVertex}
\end{figure}

For skew scattering we consider only diagrams with a single third order impurity vertex and both external current vertices renormalized by ladder vertex corrections as indicated in Fig.~\ref{fig:SigmaIa}(c). In analogy to the renormalized $v_x$-vertex in Eq.~(\ref{eq:RenVertex}) also the renormalized $v_y$-vertex can be calculated and expressed via $\Gamma_x$ and $\Gamma_y$ as
\begin{equation}
\Gamma_{v_y}=-\Gamma_y \sigma_x - \Gamma_x \sigma_y \,.
\end{equation}
Using these expressions the skew scattering diagram of Fig.~\ref{fig:SigmaIa}(c) yields
\begin{eqnarray}
\sigma_{yx}^{I(a),s}
\!\!\!&=&\!\!\frac{e^2}{2\pi} \frac{n_i V_1^3}{2\pi} 
\!\!\! \int \!\! dk k {\rm Tr} 
[\Gamma_{v_y} G^R(\epsilon_F)  \Gamma_{v_x} 
\!\!+ \Gamma_{v_y}  \Gamma_{v_x}  G^A(\epsilon_F) ] \nonumber \\
&=& \frac{e^2}{2\pi} \frac{i V_1^3}{V_0^2} {\rm Tr} [
-\Gamma_{v_y} (\Gamma \sigma_0 + \Gamma_z \sigma_z)
\Gamma_{v_x}\nonumber \\
&& \hspace{1.6cm} + \Gamma_{v_y} \Gamma_{v_x} 
                (\Gamma \sigma_0 + \Gamma_z \sigma_z)]\nonumber \\
&=& \frac{e^2}{2\pi} \frac{V_1^3}{V_0^2}i \Gamma_z {\rm Tr} [
(\Gamma_y \sigma_x + \Gamma_x \sigma_y) 
(\sigma_z (\Gamma_x \sigma_x + \Gamma_y \sigma_y)\nonumber \\
&& \hspace{3.9cm} -(\Gamma_x \sigma_x + \Gamma_y \sigma_y)
\sigma_z)]\nonumber \\
&=& \frac{e^2}{2\pi} \frac{V_1^3}{V_0^2} 4 \Gamma_z
(\Gamma_y^2-\Gamma_x^2) \,.
\end{eqnarray}
From this expression it is evident that the skew scattering contribution vanishes as soon as  $\Gamma_z=0$ implying that
the lifetimes  in both bands become equal since $\Gamma_- - \Gamma_+=  \Gamma_z (h/\lambda_- + h/\lambda_+) $ vanishes
for  $\Gamma_z=0$. Plugging in $\Gamma_x$ and $\Gamma_y$ from  Eq.~(\ref{eq:vertexcorrections}) one
finds~\cite{Borunda:2007_a} in the weak scattering limit, i.e., neglecting higher order impurity terms: 
\begin{eqnarray}
\sigma_{yx}^{I(a),s} &=& \frac{e^2}{2 \pi}
\frac{4 V_1^3 \Gamma_z \gamma_y^2}{V_0^2 (1- n_i V_0^2 I_1)^2} \\
&=& \frac{e^2}{2 \pi} \frac{V_1^3}{n_i V_0^4}
\frac{h \lambda_- \alpha^2 k_-^4}{\nu_- (3 h^2 + \lambda_-^2)^2}
\,.
\label{eq:SkewScattering}
\end{eqnarray}
It can be shown easily that considering the weak scattering limit of the full vertex shown in Fig.~\ref{fig:SkewVertex} yields exactly the same result as Eq.~(\ref{eq:SkewScattering}), i.e., to order $V_1^3/(n_i V_0^4)$ it reduces to the elementary skew scattering diagram depicted in Fig.~\ref{fig:SigmaIa}(c).

\subsection{Simple limits}
\label{sec:SimpleLimits}

\subsubsection{Both subbands occupied}
\label{sec:BothSubbands}

In the situation that both subbands are partially occupied, i.e.,
$\epsilon_F > h$, all contributions to the anomalous Hall conductivity
vanish. For $\sigma_{yx}^{II}$ this is immediately evident from
Eq.~(\ref{eq:sigma2}). For the skew scattering contribution, which   
is proportional to $\Gamma_z$ (see Eq.~(\ref{eq:SkewScattering})),
one observes easily that $\sigma_{yx}^{I(a),s}=0$ because $\Gamma_z=0$ (see
Eq.~(\ref{eq:SelfEnergy})) due to $\nu_+/\lambda_+ - \nu_-/\lambda_- =0$ 
(see Eq.~(\ref{eq:nu})).

With respect to the bare bubble and ladder diagrams we will show in
the following that they cancel mutually. For $\epsilon_F > h$ the
integrals in Eq.~(\ref{eq:integrals}) simplify to
\begin{eqnarray}
I_1 = \frac{\alpha^2 m^2 \epsilon_F}{2 \lambda_F^2 \Gamma} \,,\,\,\,\,
I_2 = - \frac{ihm}{2 \lambda_F^2} \,,\,\,\,\,
I_3=0 \,,\,\,\,\,
I_4=\frac{\alpha m}{2 \Gamma}
\end{eqnarray}
and the bare momentum integrated vertices in
Eq.~(\ref{eq:BareVertices}) are:
\begin{equation}
n_i V_0^2 \gamma_x = - \frac{\alpha h \Gamma}{\lambda_F^2} \,, \quad
n_i V_0^2 \gamma_y = \alpha \left (1- \frac{\alpha^2 m \epsilon_F}{\lambda_F^2}
\right) \,.
\end{equation}
This gives for the bare bubble in Eq.~(\ref{eq:BareBubble})
\begin{equation}
\sigma_{yx}^{I(a),b}=-\frac{e^2}{2\pi}\frac{\alpha^2 mh}{\lambda_F^2} \,.
\end{equation}
For the ladder diagrams we need also
\begin{equation}
1-n_i V_0^2 I_1 = \frac{n_i V_0^2 \gamma_y}{\alpha} \,,\quad
- i n_i V_0^2 I_2 =  \frac{n_i V_0^2 \gamma_x}{\alpha}
\end{equation}
yielding
\begin{eqnarray}
\sigma_{yx}^{I(a),l} \!\!&=&\!\!-\frac{e^2}{\pi} 
\frac{\alpha}{n_i V_0^2} 
\frac{2 \gamma_x \gamma_y^2 - \gamma_x \gamma_y^2+ \gamma_x^3}
{\gamma_x^2 + \gamma_y^2} 
= -\frac{e^2}{\pi} \frac{\alpha \gamma_x}{n_i V_0^2} \nonumber \\
&=& \frac{e^2}{2\pi}\frac{\alpha^2 mh}{\lambda_F^2} 
\end{eqnarray}
and thus
\begin{equation}
\sigma_{yx}^{I(a),b}+\sigma_{yx}^{I(a),l}=0 \,,
\end{equation}
i.e., the contribution of the bare bubble and the ladder diagrams
cancel mutually.

\subsubsection{Only majority band occupied}

In the opposite situation, where only the majority band is partially
occupied, we have $\nu_+=0$ and therefore $\Gamma_z \ne 0$. In this
case all terms contribute to the anomalous Hall conductivity. In the following we
restrict our analysis to Fermi energies $\epsilon_F > -h$, i.e., we
disregard the region of very small Fermi energies, where the valley
structure of the 
majority band becomes important (see Fig.~\ref{fig:disp}(b)) and
discuss the results in two simple limits: (i) small spin orbit
interaction: $\alpha k_F \ll h$ and (ii) small magnetization $ h \ll
\alpha k_F$. 

In the limit of small spin-orbit interaction $\alpha k_F \ll h$ the sum of bare bubble and ladder vertex  corrections becomes
\begin{equation}
\sigma_{yx}^{I(a),b}+\sigma_{yx}^{I(a),l}= \frac{e^2}{2 \pi}
\frac{(\alpha k_F)^2}{16 h \epsilon_F} 
\left( 3 \frac{\epsilon_F}{h}+1 \right)
\left( - \frac{\epsilon_F}{h}+1 \right)
\label{eq:BubbleLadderSmallAlpha}
\end{equation}
the contribution from the states of the full Fermi sea 
\begin{equation}
\sigma_{yx}^{II}=\frac{e^2}{4\pi} \frac{(\alpha k_F)^2}{2 h^2}
\label{eq:Sigma2SmallAlpha}
\end{equation}
and the skew scattering term
\begin{equation}
\sigma_{yx}^{I(a),s}=\frac{e^2}{2 \pi} 
\frac{(\alpha k_F)^2}{8 \epsilon_F n_i V_0}
\frac{V_1^3}{V_0^3}
\frac{(\epsilon_F+h)^2}{h^2} \,.
\label{eq:SkewSmallAlpha}
\end{equation}

In the opposite limit of small exchange field $h \ll \alpha k_F$,
considering first a spin-orbit interaction still smaller than the
Fermi energy $\alpha k_F \ll \epsilon_F$, we find for the sum of bare
bubble and ladder vertex corrections  
\begin{equation}
\sigma_{yx}^{I(a),b}+\sigma_{yx}^{I(a),l}= - \frac{e^2}{2 \pi}
\frac{3h \epsilon_F}{(\alpha k_F)^2}
\label{eq:BubbleLadderLargeAlpha1}
\end{equation}
and for the contribution from the states of the full Fermi sea
\begin{equation}
\sigma_{yx}^{II}=\frac{e^2}{4\pi} 
\left( 1- \frac{h}{\alpha k_F} \right)
\label{eq:Sigma2LargeAlpha1}
\end{equation}
and for the skew scattering term
\begin{equation}
\sigma_{yx}^{I(a),s}=\frac{e^2}{2 \pi}  \frac{V_1^3}{V_0^3}
\frac{2 h \epsilon_F}{n_i V_0 \alpha k_F} \,.
\label{eq:SkewLargeAlpha1}
\end{equation}

In the same limit where the exchange field is small $h \ll \alpha
k_F$, but the spin-orbit interaction is now larger than the Fermi
energy $\alpha k_F \gg \epsilon_F$ we find for the sum of bare bubble
and ladder vertex corrections  
\begin{equation}
\sigma_{yx}^{I(a),b}+\sigma_{yx}^{I(a),l}= - \frac{e^2}{2 \pi}
\frac{2 h \epsilon_F^3}{(\alpha k_F)^4}
\label{eq:BubbleLadderLargeAlpha2}
\end{equation}
and for the contribution from the states of the full Fermi sea 
\begin{equation}
\sigma_{yx}^{II}=\frac{e^2}{4\pi} 
\left( 1- \frac{2h \epsilon_F}{(\alpha k_F)^2} \right)
\label{eq:Sigma2LargeAlpha2}
\end{equation}
and the for skew scattering term
\begin{equation}
\sigma_{yx}^{I(a),s}=\frac{e^2}{2 \pi}  \frac{V_1^3}{V_0^3}
\frac{h}{n_i V_0} \,.
\label{eq:SkewLargeAlpha2}
\end{equation}

\subsection{Discussion}
\label{sec:NumericalEvaluation}

We now discuss the full evaluation of the anomalous Hall conductivity
in the limit of small spin orbit interaction $\alpha k_F \ll h$ 
and in the opposite limit of strong spin orbit interaction 
$\alpha k_F \gg h, \epsilon_F$. For the
following discussion we  will express all quantities in terms of the
exchange field $h$, which we define as $h=1$. Furthermore we will set
$m=1$, we choose $V_1=V_0$ and use an impurity concentration of
$n_i=0.1$.
\begin{figure}[h]
\begin{minipage}{.49\columnwidth}
\begin{center}
\includegraphics[width=1.3\columnwidth,clip=true]{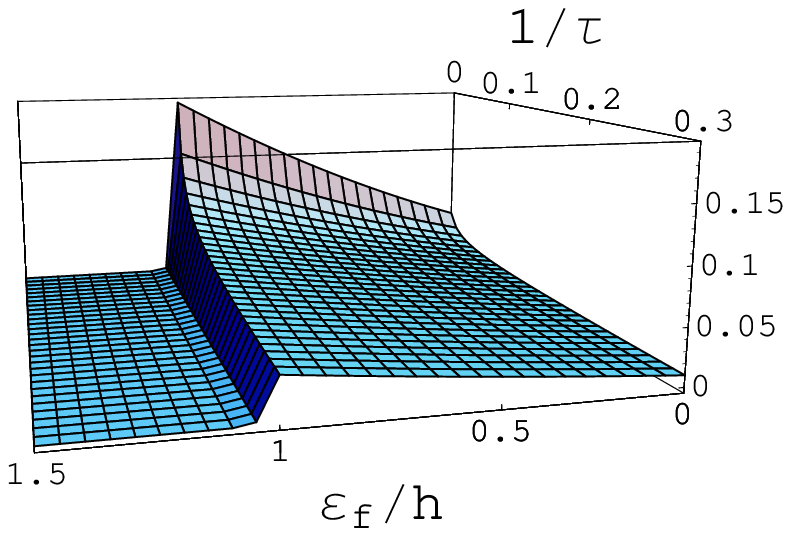}
\end{center} 
\end{minipage}
\begin{minipage}{.49\columnwidth}
\begin{center}
\includegraphics[width=1.3\columnwidth,clip=true]{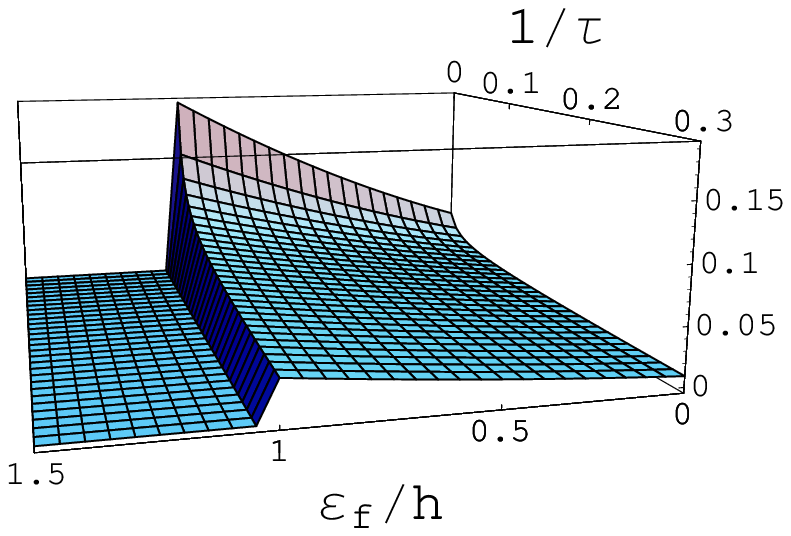}
\end{center} 
\end{minipage}\\
\begin{minipage}{.49\columnwidth}
\begin{center}\hspace*{-.5cm}
\includegraphics[width=1.3\columnwidth,clip=true]{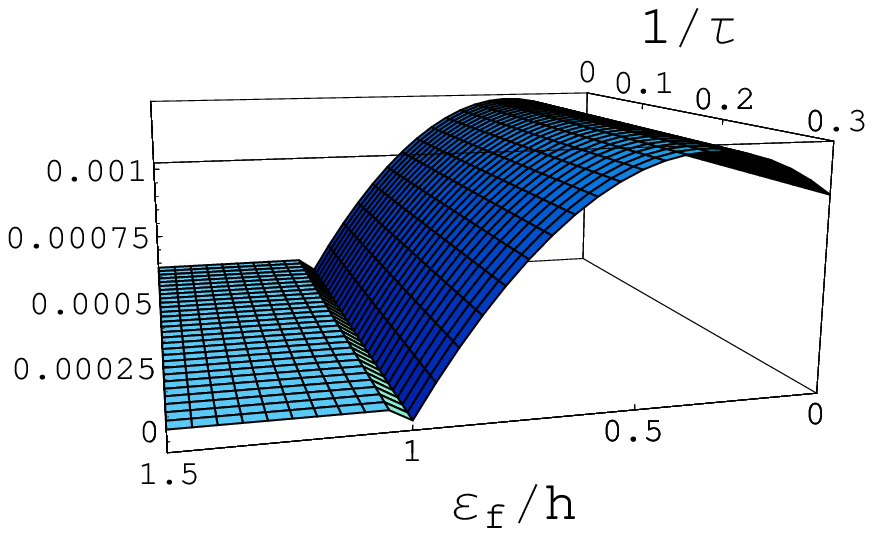}
\end{center} 
\end{minipage}
\begin{minipage}{.49\columnwidth}
\begin{center}
\includegraphics[width=1.3\columnwidth,clip=true]{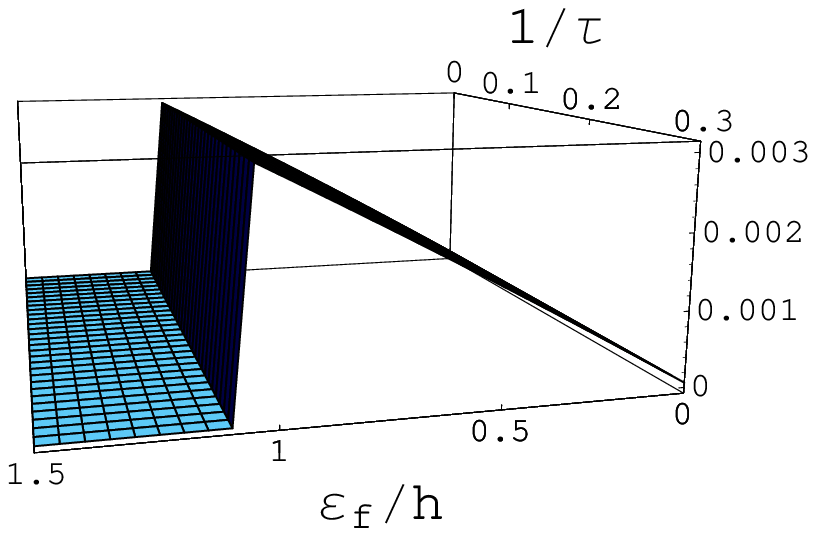}
\end{center} 
\end{minipage}
\caption{Anomalous Hall conductivity for $\alpha k_F/h=0.2$ and an impurity
  concentration of $n_i=0.1$ plotted as a function of
  $\epsilon_F/h$ (from right to left) and as a function of 
$1/\tau=n_i m V_0^2$ in units of $h$ (from back to front) where {\bf
  upper left panel:} total anomalous Hall conductivity 
  (Eq.~(\ref{eq:TotAHE})), {\bf upper right panel:} skew scattering
  contribution (Eq.~(\ref{eq:SkewScattering})), {\bf lower left panel:} bare bubble plus
  ladder vertex corrections
  (Eq.~(\ref{eq:BareBubble})+Eq.~(\ref{eq:LadderDiagrams})), {\bf lower right panel:}
  $\sigma^{II}$ (Eq.~(\ref{eq:sigma2})). All conductivities are
  plotted in units of $e^2$.}
\label{fig:AHEalpha02}
\end{figure}

In Fig.~\ref{fig:AHEalpha02} we show the anomalous Hall conductivity
for a small spin orbit interaction of $\alpha k_F/h=0.2$ as a
function of the Fermi energy $\epsilon_F/h$ and the scattering rate $1/\tau=n_i V_0^2 m$ for
an impurity concentration of $n_i=0.1$. The upper left panel shows the
total anomalous Hall conductivity, i.e., the sum of skew scattering
(upper right panel), of bare bubble and ladder diagrams (lower left
panel) and of the contribution from the whole Fermi sea (lower right
panel). Obviously all contributions to the total conductivity vanish
for $\epsilon_F > h$, i.e., when both subbands are occupied which
agrees with our analysis in Sec.~\ref{sec:BothSubbands}. 
Furthermore we observe that not only $\sigma_{yx}^{II}$ but also the
bare bubble and ladder vertex corrections
$\sigma_{yx}^{I(a),b}+\sigma_{yx}^{I(a),l}$ (see 
Eq.~(\ref{eq:BubbleLadderSmallAlpha})) are independent of impurity
scattering. Both contributions are small: 
$\sigma_{yx}^{II}$ contains a small prefactor of $(\alpha k_F/h)^2$
(see Eq.~(\ref{eq:Sigma2SmallAlpha})) and
$\sigma_{yx}^{I(a),b}+\sigma_{yx}^{I(a),l}$ a small prefactor of
$(\alpha k_F)^2/(h \epsilon_F)$ (see
Eq.~(\ref{eq:BubbleLadderSmallAlpha})). The skew scattering
contribution, on the other hand, has a prefactor of $\alpha k_F/(n_i
V_0)$ which diverges for $V_0 \to 0$, i.e, $1/\tau \to 0$ (see
Eq.~(\ref{eq:SkewSmallAlpha})) and therefore overcompensates the small
prefactor of $\alpha k_F /\epsilon_F$ (see Eq.~(\ref{eq:SkewSmallAlpha}))
when the impurity potentials $V_0$ becomes small enough.
Thus for the parameters chosen in Fig.~\ref{fig:AHEalpha02} the skew
scattering term outweighs the other contributions by orders of
magnitude and therefore the total anomalous Hall conductivity is
almost identical to the skew scattering term. It increases
quadratically with $\epsilon_F/h$ (see Eq.~(\ref{eq:SkewSmallAlpha}))
and then  vanishes suddenly for $\epsilon_F > h$.

\begin{figure}[h]
\begin{minipage}{.49\columnwidth}
\begin{center}
\includegraphics[width=1.3\columnwidth,clip=true]{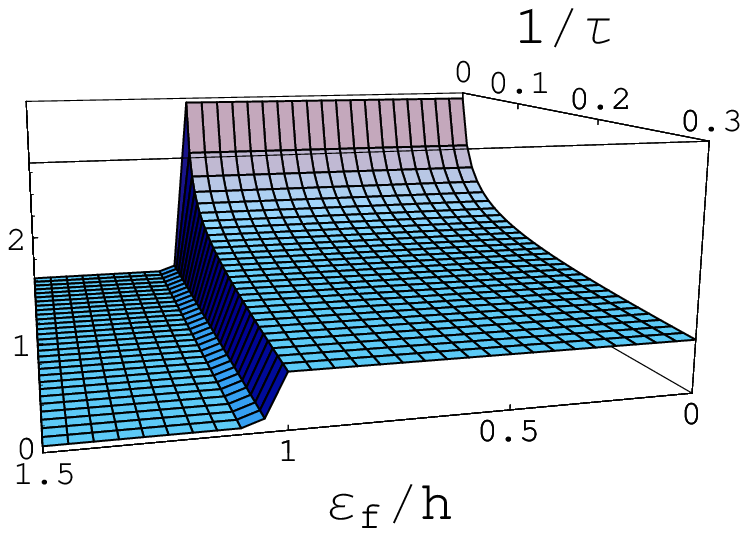}
\end{center} 
\end{minipage}
\begin{minipage}{.49\columnwidth}
\begin{center}
\includegraphics[width=1.3\columnwidth,clip=true]{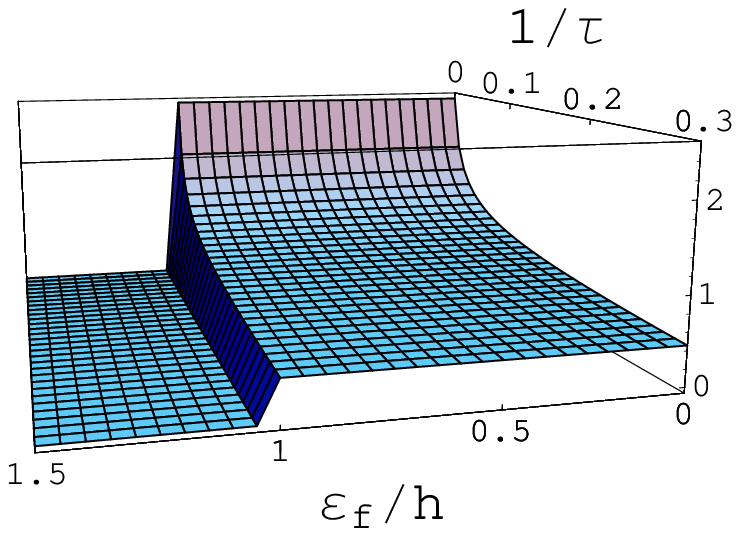}
\end{center} 
\end{minipage}\\
\begin{minipage}{.49\columnwidth}
\begin{center} \hspace*{-.5cm}
\includegraphics[width=1.3\columnwidth,clip=true]{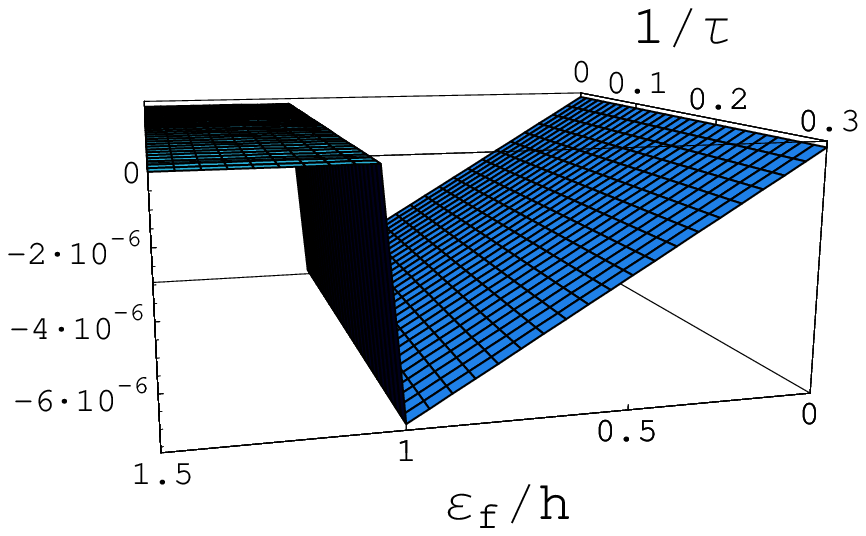}
\end{center} 
\end{minipage}
\begin{minipage}{.49\columnwidth}
\begin{center}
\includegraphics[width=1.3\columnwidth,clip=true]{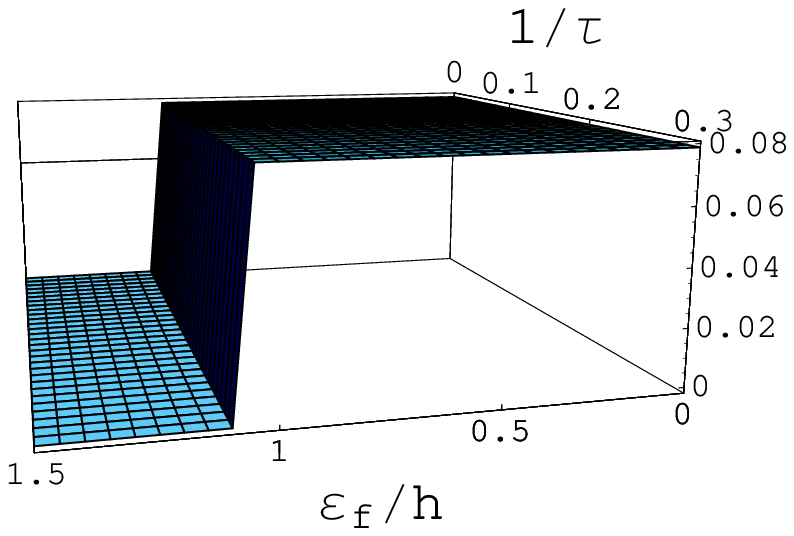}
\end{center} 
\end{minipage}
\caption{Anomalous Hall conductivity for $\alpha k_F/h=10.0$ and an impurity
  concentration of $n_i=0.1$ plotted as a function of
 $\epsilon_F/h$ (from right to left) and as a function of $1/\tau=n_i
 m V_0^2$ in units of $h$ (from back to front) where {\bf  upper left
   panel:} total anomalous Hall conductivity  
  (Eq.~(\ref{eq:TotAHE})), {\bf upper right panel:} skew scattering
  contribution (Eq.~(\ref{eq:SkewScattering})), {\bf lower left
    panel:} bare bubble plus ladder vertex corrections
  (Eq.~(\ref{eq:BareBubble})+Eq.~(\ref{eq:LadderDiagrams})), {\bf
    lower right panel:} $\sigma^{II}$ (Eq.~(\ref{eq:sigma2})). All
  conductivities are plotted in units of $e^2$.}
\label{fig:AHEalpha10}
\end{figure}

Fig.~\ref{fig:AHEalpha10} displays the anomalous Hall conductivity in
a similar way as Fig.~\ref{fig:AHEalpha02} only
for a large spin orbit interaction of $\alpha k_F/h=10$. Again,
$\sigma_{yx}^{I(a),b}+\sigma_{yx}^{I(a),l}$ turns out to be
independent of the impurity parameters and even smaller in magnitude
as before because now it is suppressed by a small prefactor of $(h
\epsilon_F^3)/(\alpha k_F)^4$ (see Eq.~(\ref{eq:BubbleLadderLargeAlpha2})). 
Analogously to the limit of small spin orbit interaction, the total
anomalous Hall conductivity is dominated by the skew scattering
contribution, which contains no small prefactor and due to the factor
of $h/(n_i V_0)$ grows rapidly for small impurity potentials $V_0 \to
0$, i.e., $1/\tau \to 0$ (see Eq.~(\ref{eq:SkewLargeAlpha2})).
In the limit of large spin-orbit interaction $\alpha k_F \gg
\epsilon_F$ the skew scattering 
and thus the total anomalous Hall conductivity is independent of the
Fermi energy $\epsilon_F$ for $\epsilon_F < h$ (see
Eq.~(\ref{eq:SkewLargeAlpha2})) and then abruptly drops to zero for
$\epsilon_F > h$. 

\section{Conclusions}
\label{sec:conclusion}

In summary, we have investigated the anomalous Hall conductivity in a
spin-polarized two-dimensional electron gas with Rashba spin-orbit
interaction in the presence of pointlike potential impurities. Our
calculations have been performed within diagrammatic
perturbation theory based on the Kubo-Streda
formula, an approach, which has previously been shown to yield  
equivalent results to the semiclassical Boltzmann
treatment.~\cite{Sinitsyn:2006_b,Borunda:2007_a}

Comparing our results with previous calculations we have been
able to sort out contradictions existing in the literature.
We have found that within the model Hamiltonian considered all contributions to the anomalous Hall
conductivity vanish as soon as the minority band becomes partially
filled, i.e., as soon as the Fermi energy becomes larger than the
internal Zeeman field.
For smaller Fermi energies all contributions are finite with
$\sigma_{yx}^{II}$, the contribution
from all states of the Fermi sea, being the smallest term at least in the
limits of weak and of strong spin orbit interaction.
The vertex corrections, which play the role of a side jump contribution,
can be of similar magnitude as the intrinsic contribution and 
turn out to be independent of the impurity concentration and impurity potential
at least in the limits of small and of strong spin orbit interaction. 
In the weak scattering limit the dominant contribution results from
skew scattering because due to its $1/(n_i V_0)$-dependence it
outweighs all other terms. Moreover, the intrinsic and the side jump
terms contain higher orders of small prefactors than the skew
scattering contribution.

\begin{acknowledgments} 
Fruitful discussions with S. Onoda and N. Nagaosa are gratefully acknowledged. This work was supported by SPP 1285 of the DFG, by ONR under Grant No. ONR-N000140610122, by the NSF under Grant no. DMR-0547875, by SWAN-NRI, by EU Grant IST-015728, by EPSRC Grant GR/S81407/01, by GACR and AVCR Grants 202/05/0575, FON/06/E002, AV0Z1010052, LC510, by the DOE under grant No. DE-AC52-06NA25396, and by the University of Kansas General Research Fund allocation No. 2302015. J.I. thanks Next Generation Super Computing Project,
Nanoscience Program, MEXT, Japan, and Grant-in-Aid for the 21st Century COE
"Frontiers of Computational Science" for financial support. Jairo Sinova is a Cottrell Scholar of the Research Foundation.
\end{acknowledgments}

\begin{appendix}
\section{Integration of $\sigma^{II}$}
\label{app:IntSigmaII}
Starting from the expression of $\sigma^{II}$ in Eq.~(\ref{eq:sigmaI_II})
one obtains after angular integration:
\begin{equation}
\sigma_{yx}^{II}= \frac{e^2}{4 \pi} \frac{1}{2\pi} \!\! \int \!\! dk k \!\!
\int_{-\infty}^{\infty} \!\!\!\! d\epsilon f(\epsilon) 
\frac{\alpha^2 h}{\lambda_k} 4 {\rm Im} \left [
G_+^R G_-^R ( G_-^R - G_+^R) \right ] .
\end{equation}
Now performing the remaining integrals in the clean limit, i.e., using
$\Gamma_+ = \Gamma_- =\delta$, yields
\begin{eqnarray}
\sigma_{yx}^{II} &=& \frac{e^2}{4 \pi} \frac{1}{2\pi} \int dk k
\int_{-\infty}^{\infty}  d\epsilon f(\epsilon) 
4 \frac{\alpha^2 h}{\lambda_k} (E_{k-}- E_{k+}) \nonumber \\
&& {\rm Im} \left [
\frac{1}{(\epsilon-E_{k+}+i\delta)^2(\epsilon -E_{k-}+i\delta)^2} \right ]
\nonumber \\
&=& \frac{e^2}{4 \pi} \frac{1}{2\pi} \int dk k
4 \frac{\alpha^2 h}{\lambda_k} 
\frac{1}{E_{k-}- E_{k+}} \\
&&\bigl \{ \pi \delta(E_{k+} -\epsilon_F)
        +\pi \delta(E_{k-} -\epsilon_F)  \nonumber \\
&&
-\frac{2}{E_{k+}- E_{k-}} {\rm Im} \left [
\ln (E_{k+} -\epsilon_F - i \delta)\right. \nonumber \\
&&\left. \hspace{2.2cm}
 -\ln (E_{k-} -\epsilon_F - i \delta) \right ] \bigr \} .
\nonumber
\end{eqnarray}
Substituting $E_{k+}- E_{k-}=2 \lambda_k$ and using
\begin{eqnarray} 
&& \int dk \frac{k}{\lambda_k^2}  \pi \delta (E_{k\pm} -\epsilon_F)
\nonumber \\
&&= \int_{\pm h}^{\infty} d E_{k\pm} 
\frac{m}{\lambda_k | \lambda_k \pm\alpha^2 m|}
\pi \delta (E_{k\pm} -\epsilon_F) \nonumber \\
&&= \frac{\pi m}{\lambda_\pm | \lambda_\pm \pm \alpha^2 m|}
\Theta(\epsilon_F-E_\pm^{\rm min})
\end{eqnarray}
and
\begin{eqnarray}
&&\int dk \frac{k}{\lambda_k^3} \ln (E_{k\pm}-\epsilon_F - i\delta)
= \left [ -\frac{\ln (E_{k\pm}-\epsilon_F - i\delta)}{\alpha^2
    \lambda_k} \right ]_0^\infty \nonumber \\
&&+ \int_0^\infty \frac{dk}{\alpha^2 \lambda_k} 
  \frac{1}{E_{k\pm}-\epsilon_F -i \delta} 
\frac{d E_{k \pm}}{dk}
\end{eqnarray}
and
\begin{eqnarray}
&&-  \left [ -\frac{\ln (E_{k+}-\epsilon_F - i\delta)}{\alpha^2
    \lambda_k} \right ]_0^\infty
+  \left [ -\frac{\ln (E_{k-}-\epsilon_F - i\delta)}{\alpha^2
    \lambda_k} \right ]_0^\infty \nonumber \\
&&= - \frac{i \pi}{\alpha^2 h} \Theta (h-\epsilon_F)
\end{eqnarray}
$\sigma^{II}$ simplifies to
\begin{eqnarray}
\sigma_{yx}^{II} &=& - \frac{e^2}{4 \pi} h
\left ( \frac{1}{m} \frac{1}{\lambda_- -\alpha^2 m} 
      - \frac{1}{m} \frac{1}{\lambda_+ +\alpha^2 m} \Theta(\epsilon_F- h)
\right. \nonumber \\
&& \left. \qquad\qquad -\frac{1}{h} \Theta(h-\epsilon_F) \right)
\\
&=& \frac{e^2}{4 \pi} \left ( 1-
\frac{h}{\sqrt{h^2 + 2 \alpha^2 m \epsilon_F +(\alpha^2m)^2}} \right)
\Theta(h-\epsilon_F) .
\nonumber
\end{eqnarray}

\section{Integrals in the weak scattering limit}
\label{app:Integrals}

In the weak scattering limit ($\Gamma, \Gamma_z$ small) the 
integrals over two Greens functions simplify to:
\begin{eqnarray}
&&\frac{1}{2 \pi} \int dk k f(k) G_+^R(k) G_+^A(k)  \\
&&=\frac{1}{2 \pi} \int dk k f(k) 
\frac{1}{\epsilon_F - E_{k+}+ i \Gamma_+}
\frac{1}{\epsilon_F - E_{k+}- i \Gamma_+} \nonumber \\
&&= \frac{1}{2 \pi} \int d E_{k+} \nu_+ f(k(E_{k+})) \frac{1}{\Gamma_+}
\frac{\Gamma_+}{(E_{k+}^2-\epsilon_F^2)^2 + \Gamma_+^2} \nonumber \\
&&\approx \frac{\nu_+ f(k_+)}{2 \Gamma_+}\nonumber \\
&&\frac{1}{2 \pi} \int dk k f(k) G_-^R(k) G_-^A(k) 
\approx \frac{\nu_- f(k_-)}{2 \Gamma_-} \,\,\, {\rm analogously}
\nonumber 
\end{eqnarray}
and
\begin{eqnarray}
&&\frac{1}{2 \pi} \int dk k f(k) G_+^R(k) G_-^A(k) \\
&&=   \frac{1}{2 \pi} \int dk k f(k) 
\frac{1}{\epsilon_F - E_{k+}+ i \Gamma_+}
\frac{1}{\epsilon_F - E_{k-}- i \Gamma_-} \nonumber \\
&&\approx \frac{1}{2 \pi}  \int dk k f(k) 
\left( \frac{1}{\epsilon_F - E_{k+}}-
       i \pi \delta (\epsilon_F - E_{k+}) \right)\nonumber \\
&&\hspace{2.6cm} \left( \frac{1}{\epsilon_F - E_{k-}}+
       i \pi \delta (\epsilon_F - E_{k-}) \right) \nonumber \\
&&\approx \frac{1}{2 \pi} \int dk k f(k) 
\frac{1}{\epsilon_F - \epsilon_k - \lambda_k}
\frac{1}{\epsilon_F - \epsilon_k + \lambda_k} \nonumber \\
&&+\frac{i}{2} \int dk k f(k) \left( 
\delta(\epsilon_F-\epsilon_k + \lambda_k) \frac{1}{\epsilon_F -
  \epsilon_k - \lambda_k} \right. \nonumber  \\
&&\left. \hspace{2.3cm}
-\delta(\epsilon_F-\epsilon_k - \lambda_k) \frac{1}{\epsilon_F - \epsilon_k + \lambda_k}
\right) \nonumber
\end{eqnarray}
yielding
\begin{eqnarray}
&&\frac{1}{2 \pi} \int  dk k f(k) (G_+^R(k) G_-^A(k)- G_-^R(k)
G_+^A(k)) \nonumber \\
&&\approx i  \int dE_{k-}  
\frac{\nu_- f(k(E_{k-})) \delta( \epsilon_F - E_{k-})}
{\epsilon_F - E_{k-} - 2 \lambda_{k(E_{k-})}} \nonumber \\
&& -i \int  dE_{k+}  
\frac{\nu_+ f(k(E_{k+})) \delta( \epsilon_F - E_{k+})}
{\epsilon_F - E_{k+} + 2 \lambda_{k(E_{k+})}} \nonumber \\
&& = - \frac{i}{2} \left( \frac{\nu_+ f(k_+)}{\lambda_+}
                       +\frac{\nu_- f(k_-)}{\lambda_-} \right) .
\end{eqnarray}

Now we find for the integrals $I_1$, $I_2$, $I_3$ and $I_4$
in the weak scattering limit: 
\begin{eqnarray}
I_1 \!\!&=&\!\! \frac{1}{2\pi} \int dk k \left (G_0^R G_0^A - G_z^R G_z^A
\right)  \\
\!\!&=&\!\! \frac{1}{4} \frac{1}{2\pi} \int dk k 
\left ( G_+^R G_+^A + G_-^R G_-^A + G_+^R G_-^A + G_-^R G_+^A
\right.\nonumber \\
&& \!\!\!\!\!\!\left. - \frac{\lambda_k^2 (h^2 + \Gamma_z^2)}{\lambda_k^4 + h^2 \Gamma_z^2}
\left( G_+^R G_+^A + G_-^R G_-^A - G_+^R G_-^A - G_-^R G_+^A \right)\!\right)
\nonumber \\
&\approx& \frac{1}{4} \frac{1}{2\pi} \int dk k  
\left( 1- \frac{h^2}{\lambda_k^2} \right) 
\left( G_+^R G_+^A + G_-^R G_-^A \right)
\nonumber \\
&\approx& \frac{1}{8} \left (
          \left(1-\frac{h^2}{\lambda_+^2} \right) \frac{\nu_+}{\Gamma_+}
          +\left(1-\frac{h^2}{\lambda_-^2} \right) \frac{\nu_-}{\Gamma_-} 
\right ) \nonumber \\
I_2 \!\! &=& \!\!\! \frac{1}{2\pi} \int dk k \left (G_0^R G_z^A - G_z^R G_0^A \right)
\\
\!\!&=& \!\!\! - \frac{1}{2} \frac{1}{2\pi} \int \!\! dk k 
\frac{\lambda_k}{\lambda_k^4 + \Gamma_z^2 h^2}
\left( h (\lambda_k^2 \!+\! \Gamma_z^2)( G_-^R G_+^A \!-\! G_+^R G_-^A)
\right.
\nonumber \\
&&\hspace{2.8cm} + \left. i \Gamma_z (h^2 \!-\!\lambda_k^2) (G_+^R G_+^A \!-\! G_-^R G_-^A) \right)
\nonumber \\
&\approx& \!\!\!- \frac{1}{2} \frac{1}{2\pi} \int \!\! dk k 
\frac{1}{\lambda_k^3}
\left( h \lambda_k^2 (- G_+^R G_-^A + G_-^R G_+^A) \right. \nonumber \\
&&\hspace{2.1cm} \left. + i \Gamma_z (h^2 -\lambda_k^2) (G_+^R G_+^A - G_-^R G_-^A) \right)
\nonumber \\
&\approx& \!\!\!- \frac{i}{4} \! \left ( \!
\frac{\nu_+ h}{\lambda_+^2} \!+\! \frac{\nu_- h}{\lambda_-^2}
\!+\! \frac{\Gamma_z}{\Gamma_+} \frac{\nu_+ (h^2 \!-\!\lambda_+^2)}{\lambda_+^3}  
\!-\! \frac{\Gamma_z}{\Gamma_-} \frac{\nu_- (h^2\!-\!\lambda_-^2)}{\lambda_-^3}  
\! \right ) \nonumber \\
I_3 \!\!&=& \!\!\frac{1}{2\pi} \int dk \frac{k^2}{2m} 
\left (\tilde G_y^R G_z^A - G_z^R \tilde G_y^A \right)
\nonumber \\
\!\!&=&\!\! -\frac{i}{2} \frac{1}{2\pi} \int dk k \frac{k^2}{2m} 
\frac{\alpha \Gamma_z \lambda_k^2}{\lambda_k^4 + \Gamma_z^2 h^2}
( G_+^R G_+^A + G_-^R G_-^A \\
&& \hspace{4.2cm} - G_+^R G_-^A - G_-^R G_+^A)
\nonumber \\
&\approx& \!\!-\frac{i}{2} \frac{1}{2\pi} \int dk k \frac{k^2}{2m} 
\frac{\alpha \Gamma_z}{\lambda_k^2}
( G_+^R G_+^A + G_-^R G_-^A)
\nonumber \\
&\approx& \!\!-\frac{i}{4} \alpha \Gamma_z \left (
\epsilon_F \left( \frac{\nu_+}{\Gamma_+ \lambda_+^2}
                 +\frac{\nu_-}{\Gamma_- \lambda_-^2} \right )
-\frac{\nu_+}{\Gamma_+ \lambda_+}+\frac{\nu_-}{\Gamma_- \lambda_-}  
\right ) \nonumber \\
I_4 \!\!&=& \!\!\frac{1}{2\pi} \int dk \frac{k^2}{2m} 
\left (G_0^R \tilde G_y^A + \tilde G_y^R G_0^A \right)
\\
\!\!&=& \!\!- \frac{1}{2} \frac{1}{2\pi} \int dk k \frac{k^2}{2m}
\frac{\alpha \lambda_k}{\lambda_k^4 + \Gamma_z^2 h^2}
\left( \lambda_k^2 (G_+^R G_+^A - G_-^R G_-^A) \right. \nonumber \\
&&\hspace{3.8cm} \left.+ i \Gamma_z h (G_-^R G_+^A-G_+^R G_-^A ) \right)
\nonumber \\
&\approx& \!\!- \frac{1}{2} \frac{1}{2\pi} \int dk k \frac{k^2}{2m}
\frac{\alpha}{\lambda_k} (G_+^R G_+^A - G_-^R G_-^A) 
\nonumber \\
&\approx&\!\! - \frac{1}{4} \alpha \left (
\epsilon_F \left( \frac{\nu_+}{\Gamma_+ \lambda_+}
                 -\frac{\nu_-}{\Gamma_- \lambda_-} \right )
-\left( \frac{\nu_+}{\Gamma_+}+\frac{\nu_-}{\Gamma_-} \right) 
\right ) . \nonumber
\end{eqnarray}

\end{appendix}


\begin{thebibliography}{24}
\expandafter\ifx\csname natexlab\endcsname\relax\def\natexlab#1{#1}\fi
\expandafter\ifx\csname bibnamefont\endcsname\relax
  \def\bibnamefont#1{#1}\fi
\expandafter\ifx\csname bibfnamefont\endcsname\relax
  \def\bibfnamefont#1{#1}\fi
\expandafter\ifx\csname citenamefont\endcsname\relax
  \def\citenamefont#1{#1}\fi
\expandafter\ifx\csname url\endcsname\relax
  \def\url#1{\texttt{#1}}\fi
\expandafter\ifx\csname urlprefix\endcsname\relax\def\urlprefix{URL }\fi
\providecommand{\bibinfo}[2]{#2}
\providecommand{\eprint}[2][]{\url{#2}}

\bibitem{Hall} E.~H.~Hall, Philos. Mag. {\bf 10}, 301 (1880);
  Philos. Mag. {\bf 12}, 157 (1881).

\bibitem[{\citenamefont{Sinova et~al.}(2004)\citenamefont{Sinova, Jungwirth,
  and Cerne}}]{Sinova:2004_c}
\bibinfo{author}{\bibfnamefont{J.}~\bibnamefont{Sinova}},
  \bibinfo{author}{\bibfnamefont{T.}~\bibnamefont{Jungwirth}}, \bibnamefont{and}
  \bibinfo{author}{\bibfnamefont{J.} \bibnamefont{Cerne}},
  \bibinfo{journal}{Int. J. Mod. Phys.} \textbf{\bibinfo{volume}{B 18}},
  \bibinfo{pages}{1083} (\bibinfo{year}{2004}).

\bibitem[{\citenamefont{Karplus and Luttinger}(1954)}]{Karplus:1954_a}
\bibinfo{author}{\bibfnamefont{R.}~\bibnamefont{Karplus}} \bibnamefont{and}
  \bibinfo{author}{\bibfnamefont{J.~M.} \bibnamefont{Luttinger}},
  \bibinfo{journal}{Phys. Rev.} \textbf{\bibinfo{volume}{95}},
  \bibinfo{pages}{1154} (\bibinfo{year}{1954}).

\bibitem[{\citenamefont{Smit}(1955)}]{Smit:1955_a}
\bibinfo{author}{\bibfnamefont{J.}~\bibnamefont{Smit}},
  \bibinfo{journal}{Physica} \textbf{\bibinfo{volume}{21}},
  \bibinfo{pages}{877} (\bibinfo{year}{1955}).

\bibitem{note_skew} Note that the origin of the asymmetry of this scattering arises from the spin-orbit coupling present in the Bloch states and not from the very weak spin-orbit coupling contribution of the disorder potential as noted originally by Smit. When projecting a multi-band system to an effective conduction band system one can obtain a term that looks as if it arises from such a spin-orbit coupling part of the disorder potential but it truly originates from spin-orbit coupling induced by the valence band states and the normal disorder that is felt by them.

\bibitem[{\citenamefont{Berger}(1970)}]{Berger:1970_a}
\bibinfo{author}{\bibfnamefont{L.}~\bibnamefont{Berger}},
  \bibinfo{journal}{Phys. Rev.} \textbf{\bibinfo{volume}{B 2}},
  \bibinfo{pages}{4559} (\bibinfo{year}{1970}).

\bibitem[{\citenamefont{Nozieres and Lewiner}(1973)}]{Nozieres:1973_a}
\bibinfo{author}{\bibfnamefont{P.}~\bibnamefont{Nozieres}} \bibnamefont{and}
  \bibinfo{author}{\bibfnamefont{C.} \bibnamefont{Lewiner}},
  \bibinfo{journal}{Le Journal de Physique} \textbf{\bibinfo{volume}{34}},
  \bibinfo{pages}{901} (\bibinfo{year}{1973}).

\bibitem[{\citenamefont{Kohn and Luttinger}(1957)}]{Kohn:1957_a}
\bibinfo{author}{\bibfnamefont{W.}~\bibnamefont{Kohn}} \bibnamefont{and}
  \bibinfo{author}{\bibfnamefont{J.~M.} \bibnamefont{Luttinger}},
  \bibinfo{journal}{Phys. Rev.} \textbf{\bibinfo{volume}{108}},
  \bibinfo{pages}{590} (\bibinfo{year}{1957}).

\bibitem[{\citenamefont{Luttinger}(1958)}]{Luttinger:1958_a}
 \bibinfo{author}{\bibfnamefont{J.~M.} \bibnamefont{Luttinger}},
  \bibinfo{journal}{Phys. Rev.} \textbf{\bibinfo{volume}{112}},
  \bibinfo{pages}{739} (\bibinfo{year}{1958}).

\bibitem[{\citenamefont{Sinitsyn et~al.}(2006)\citenamefont{Sinitsyn,
  MacDonald, Jungwirth, Dugaev, and Sinova}}]{Sinitsyn:2006_b}
\bibinfo{author}{\bibfnamefont{N.~A.} \bibnamefont{Sinitsyn}},
  \bibinfo{author}{\bibfnamefont{A.~H.} \bibnamefont{MacDonald}},
  \bibinfo{author}{\bibfnamefont{T.}~\bibnamefont{Jungwirth}},
  \bibinfo{author}{\bibfnamefont{V.~K.} \bibnamefont{Dugaev}},
  \bibnamefont{and} \bibinfo{author}{\bibfnamefont{J.}~\bibnamefont{Sinova}}
  Phys. Rev. B {\bf 75}, 045315 (\bibinfo{year}{2007}).

\bibitem[{\citenamefont{Culcer}(2003)}]{Culcer:2003_a}
\bibinfo{author}{\bibfnamefont{D.}~\bibnamefont{Culcer}},
  \bibinfo{author}{\bibfnamefont{A.~H.} \bibnamefont{MacDonald}}, \bibnamefont{and}
  \bibinfo{author}{\bibfnamefont{Q.} \bibnamefont{Niu}},
  \bibinfo{journal}{Phys. Rev.} \textbf{\bibinfo{volume}{B 68}},
  \bibinfo{pages}{045327} (\bibinfo{year}{2003}).

\bibitem[{\citenamefont{Dugaev et~al.}(2005)\citenamefont{Dugaev, Bruno,
  Taillefumier, Canals, and Lacroix}}]{Dugaev:2005_a}
\bibinfo{author}{\bibfnamefont{V.~K.} \bibnamefont{Dugaev}}, 
  \bibinfo{author}{\bibfnamefont{P.}~\bibnamefont{Bruno}},
  \bibinfo{author}{\bibfnamefont{M.}~\bibnamefont{Taillefumier}},
  \bibinfo{author}{\bibfnamefont{B.}~\bibnamefont{Canals}}, \bibnamefont{and}
  \bibinfo{author}{\bibfnamefont{C.}~\bibnamefont{Lacroix}},
  \bibinfo{journal}{Phys. Rev.} \textbf{\bibinfo{volume}{B 71}},
  \bibinfo{pages}{224423} (\bibinfo{year}{2005}).

\bibitem[{\citenamefont{Sinitsyn et~al.}(2005)\citenamefont{Sinitsyn, Niu,
  Sinova, and Nomura}}]{Sinitsyn:2005_a}
\bibinfo{author}{\bibfnamefont{N.~A.} \bibnamefont{Sinitsyn}},
  \bibinfo{author}{\bibfnamefont{Q.}~\bibnamefont{Niu}},
  \bibinfo{author}{\bibfnamefont{J.}~\bibnamefont{Sinova}}, \bibnamefont{and}
  \bibinfo{author}{\bibfnamefont{K.}~\bibnamefont{Nomura}},
  \bibinfo{journal}{Phys. Rev.} \textbf{\bibinfo{volume}{B 72}},
  \bibinfo{pages}{045346} (\bibinfo{year}{2005}).

\bibitem[{\citenamefont{Liu and Lei}(2005)}]{Liu:2005_c}
\bibinfo{author}{\bibfnamefont{S.~Y.} \bibnamefont{Liu}} \bibnamefont{and}
  \bibinfo{author}{\bibfnamefont{X.~L.} \bibnamefont{Lei}},
  \bibinfo{journal}{Phys. Rev.} \textbf{\bibinfo{volume}{B 72}},
  \bibinfo{pages}{195329} (\bibinfo{year}{2005}).

\bibitem[{\citenamefont{Liu and Lei}(2005)}]{Liu:2006_a}
\bibinfo{author}{\bibfnamefont{S.~Y.} \bibnamefont{Liu}},
 \bibinfo{author}{\bibfnamefont{N.~J.~M.} \bibnamefont{Horing}}, \bibnamefont{and}
  \bibinfo{author}{\bibfnamefont{X.~L.} \bibnamefont{Lei}},
  \bibinfo{journal}{Phys. Rev.} \textbf{\bibinfo{volume}{B 74}},
  \bibinfo{pages}{165316} (\bibinfo{year}{2006}).

\bibitem[{\citenamefont{ichiro Inoue et~al.}(2006)\citenamefont{ichiro Inoue,
  Kato, Ishikawa, Itoh, Bauer, and Molenkamp}}]{Inoue:2006_a}
\bibinfo{author}{\bibfnamefont{J.}~\bibnamefont{Inoue}},
  \bibinfo{author}{\bibfnamefont{T.}~\bibnamefont{Kato}},
  \bibinfo{author}{\bibfnamefont{Y.}~\bibnamefont{Ishikawa}},
  \bibinfo{author}{\bibfnamefont{H.}~\bibnamefont{Itoh}},
  \bibinfo{author}{\bibfnamefont{G.~E.~W.} \bibnamefont{Bauer}},
  \bibnamefont{and} \bibinfo{author}{\bibfnamefont{L.~W.}
  \bibnamefont{Molenkamp}}, 
\bibinfo{journal}{Phys. Rev. Lett.}
  \textbf{\bibinfo{volume}{97}}, \bibinfo{pages}{046604}
  (\bibinfo{year}{2006}).

\bibitem[{\citenamefont{Onoda et~al.}(2006)\citenamefont{Onoda, Sugimoto, and
  Nagaosa}}]{Onoda:2006_a}
\bibinfo{author}{\bibfnamefont{S.}~\bibnamefont{Onoda}},
  \bibinfo{author}{\bibfnamefont{N.}~\bibnamefont{Sugimoto}}, \bibnamefont{and}
  \bibinfo{author}{\bibfnamefont{N.}~\bibnamefont{Nagaosa}},
 Phys. Rev. Lett. {\bf 97}, 126602 (2006).

\bibitem{Borunda:2007_a} M. F. Borunda, T. S. Nunner, T. Luck, 
N. A. Sinitsyn, C. Timm, J. Wunderlichl, T. Jungwirth, A. H. MacDonald, and J. Sinova,
  cond-mat/0702289 (\bibinfo{year}{2007}).


\bibitem{Sinitsyn:2006_a} N. A. Sinitsyn, Q. Niu, and A. H. MacDonald
Phys. Rev. B {\bf 73}, 075318 (2006).

\bibitem[{\citenamefont{Streda}(1982)\citenamefont{Streda}}]{Streda:1982_a}
\bibinfo{author}{\bibfnamefont{P.} \bibnamefont{Streda}},
\bibinfo{journal}{J. Phys.}
  \textbf{\bibinfo{volume}{C 15}}, \bibinfo{pages}{L717} (\bibinfo{year}{1982}).

\bibitem{Kovalev:2007_b} A. A. Kovalev {\it et al.}, unpublished.

\end{thebibliography}

\end{document}